\documentclass[sigplan,screen]{acmart}

\usepackage{subcaption}
\usepackage{booktabs}
\usepackage{multirow}

\AtBeginDocument{%
  }

\setcopyright{acmlicensed}
\copyrightyear{2018}
\acmYear{2018}
\acmDOI{XXXXXXX.XXXXXXX}
\acmConference[Conference acronym 'XX]{Make sure to enter the correct
  conference title from your rights confirmation email}{June 03--05,
  2018}{Woodstock, NY}
\acmISBN{978-1-4503-XXXX-X/2018/06}

\begin{document}

\title{StreamDQ: Near-Memory Weight DeQuantization in Custom HBM for Scalable AI Inference Acceleration}


\author{Minki Jeong}
\affiliation{%
  \institution{SK Hynix}
  \city{Icheon}
  \country{South Korea}}
\email{minki.jeong@sk.com}

\author{Daegun Yoon}
\affiliation{%
  \institution{SK Hynix}
  \city{Icheon}
  \country{South Korea}}
\email{daegun.yoon@sk.com}

\author{Soohong Ahn}
\affiliation{%
  \institution{SK Hynix}
  \city{Icheon}
  \country{South Korea}}
\email{soohong.ahn@sk.com}

\author{Seungyong Lee}
\affiliation{%
  \institution{SK Hynix}
  \city{Icheon}
  \country{South Korea}}
\email{seungyong.lee@sk.com}

\author{Nameun Kang}
\affiliation{%
  \institution{SK Hynix}
  \city{Icheon}
  \country{South Korea}}
\email{nameun.kang@sk.com}

\author{Hyeonseok Ju}
\affiliation{%
  \institution{SK Hynix}
  \city{Icheon}
  \country{South Korea}}
\email{hyeonseok.ju@sk.com}

\author{Ieryung Park}
\affiliation{%
  \institution{SK Hynix}
  \city{Icheon}
  \country{South Korea}}
\email{ieryung.park@sk.com}

\author{Joonseop Sim}
\affiliation{%
  \institution{SK Hynix}
  \city{Icheon}
  \country{South Korea}}
\email{joonseop.sim@sk.com}

\author{Youngpyo Joo}
\affiliation{%
  \institution{SK Hynix}
  \city{Icheon}
  \country{South Korea}}
\email{youngpyo.joo@sk.com}

\author{Hoshik Kim}
\affiliation{%
  \institution{SK Hynix}
  \city{Icheon}
  \country{South Korea}}
\email{hoshik.kim@sk.com}

\renewcommand{\shortauthors}{Trovato et al.}

\begin{abstract}
As large language models (LLMs) scale, their memory and computation demands have grown substantially, making weight-only quantization a widely adopted technique for reducing model size with minimal accuracy loss. However, on current GPUs, CUDA-core-based dequantization introduces substantial instruction overhead, on-chip traffic, and pipeline stalls, making it a major bottleneck for high-throughput, cloud-scale LLM serving. To address these limitations, we propose StreamDQ, a lightweight architectural enhancement that enables on-the-fly dequantization in the memory subsystem for high-throughput, large-batch LLM inference. StreamDQ integrates compact DeQuantization Blocks (DQBs) into the base die of high-bandwidth memory (HBM) and performs inline dequantization on standard memory loads. A lightweight sideband tag on each memory read request selects the dequantization mode while preserving conventional load semantics.
By relocating dequantization to the memory side, StreamDQ eliminates GPU-side CUDA-core-based dequantization, thereby reducing on-chip traffic on the GPU and avoiding extra HBM write-back and reload of dequantized weights at large batch sizes. Our evaluation shows that StreamDQ achieves up to 7.08$\times$ speedup and 90.23\% lower energy for mixed-precision GEMM, with only 0.127\,mm$^2$ area and 0.355\,W power overhead per DQB in a 12\,nm CMOS process. For end-to-end LLM inference, StreamDQ reduces latency by up to 54.68\% and improves decode throughput by up to 2.20$\times$.
\end{abstract}


\maketitle

\section{Introduction}

Large Language Models (LLMs) have achieved remarkable success across a wide range of natural language processing tasks~\cite{touvron2023llama,zhang2022opt,liu2024deepseek}, driven largely by empirical scaling laws that correlate model size with performance~\cite{kaplan2020scaling,bahri2024explaining}. However, this rapid growth substantially increases computation and memory demands during inference, where billions of parameters must be fetched and processed in real time. As a result, data movement becomes a major source of latency, energy consumption, and deployment cost, making efficient LLM serving increasingly challenging~\cite{kwon2023efficient}.

Quantization~\cite{dai2021vs,shao2023omniquant,xia2024quant,guo2022ant,chee2023quip,guo2023olive,lin2024awq} is a widely adopted technique for mitigating these costs by representing model weights and activations in reduced precision. It maps high-precision floating-point data (e.g., FP32) into compact integer or lower-precision floating-point formats (e.g., INT4 or FP8), and the corresponding quantization and dequantization can be expressed as
\begin{equation}
    Quant(x) = \mathrm{round}\Big(\frac{x}{s} + z\Big), \quad Dequant(x) = (x - z) \cdot s
    \label{equation_quant}
\end{equation}
where \(s\) and \(z\) denote the scaling factor ($S$) and zero-point offset ($Z$), respectively~\cite{jacob2018quantization}. By reducing memory footprint and improving bandwidth utilization, quantization is beneficial for transformer-based LLMs~\cite{vaswani2017attention}, where linear layers dominate both computation and memory traffic due to extensive general matrix multiplication (GEMM) operations~\cite{vaswani2017attention,cho2021accelerating,jang2024figna,zhu2025nanoflow}. Thus, improving GEMM efficiency is critical for reducing inference latency and energy.

Existing quantization techniques can be broadly categorized into weight-activation quantization and weight-only quantization. Weight-activation quantization~\cite{dettmers2022gpt3,xiao2023smoothquant,wei2022outlier,yao2022zeroquant,zhao2024atom,lin2024qserve} jointly reduces the precision of both weights and activations (e.g., W8A8 using INT8 weights and activations), enabling fully low-precision GEMM execution on tensor cores. However, activation quantization is highly sensitive to outlier values, often causing notable accuracy degradation. LLM.int8~\cite{dettmers2022gpt3} addresses this by preserving activation outliers in FP16, while SmoothQuant~\cite{xiao2023smoothquant} shifts the burden of outliers from activations to weights to enable more stable quantization. Although effective, these methods introduce non-trivial system complexity and compile-time overhead, limiting their scalability and practicality for cloud-scale LLM serving.

\begin{figure}
\centering
\includegraphics[width=\columnwidth]{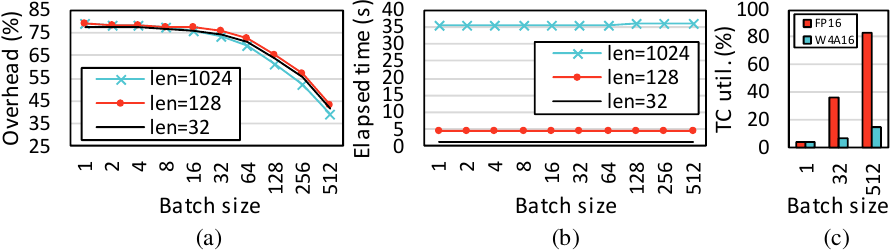}
\vspace{-2em}
\caption{
DeQuantization (DQ) overhead in LLaMA-3.1-8B-Instruct inference (W4A16) and tensor-core utilization:
(a) DQ overhead as a fraction of inference time ("len = x" denotes both the input and output sequence lengths).
(b) Absolute DQ elapsed time.
(c) Tensor-core utilization of FP16 fpGEMM and W4A16 mpGEMM with a 4096$\times$4096 weight matrix.
}
\label{fig:dq-overhead}
\vspace{-1.5em}
\end{figure}

In contrast, weight-only quantization~\cite{gptq,lee2024owq,lin2024awq,kim2023squeezellm,fang2025anda,jeon2022mr,park2022lut,jang2024figna} retains high-precision activations while quantizing only static weights offline (e.g., W4A16), and remains highly effective in practice because model weights dominate both the memory footprint and bandwidth demand of LLM inference. This approach provides substantial memory and bandwidth savings with minimal accuracy loss, but it exposes a fundamental architectural inefficiency: current GPUs generally lack native support for weight-only mixed-precision GEMM (mpGEMM), such as INT4$\times$FP16. Consequently, dequantization is typically executed on CUDA cores prior to GEMM, introducing significant instruction overhead, on-chip data movement, and pipeline stalls that become more pronounced under high-throughput workloads.
As illustrated in Fig.~\ref{fig:dq-overhead}-(a), dequantization accounts for up to 40--80\% of total inference latency across batch sizes in LLaMA-3.1-8B-Instruct (W4A16). Although its absolute execution time remains relatively stable, as shown in Fig.~\ref{fig:dq-overhead}-(b), its latency contribution remains substantial, making dequantization a major performance bottleneck.

\begin{figure}[t]
\centering
\includegraphics[width=0.95\columnwidth]{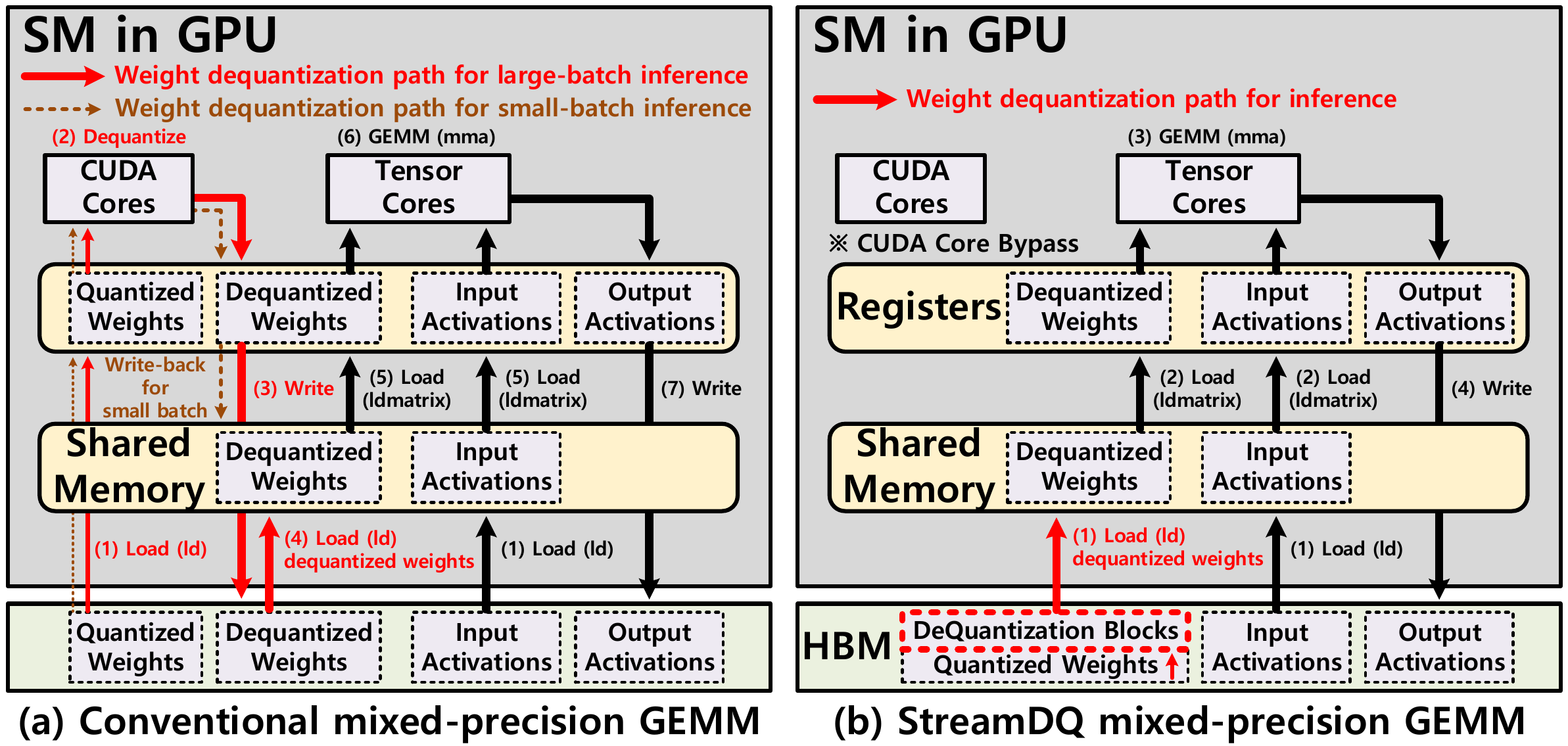}
\caption{
Overview of mpGEMM under weight-only quantization, comparing (a) the conventional GPU-based workflow and (b) the proposed StreamDQ-based workflow.
}
\label{kernel}
\vspace{-1.5em}
\end{figure}

To mitigate redundant off-chip memory accesses, frameworks such as vLLM~\cite{kwon2023efficient} fuse dequantization and GEMM into a single kernel using shared memory.
While this fusion reduces HBM traffic, it introduces a compute imbalance: tensor-core GEMM must wait for CUDA-core dequantization to complete, lowering overall tensor-core utilization relative to FP16 full-precision GEMM (fpGEMM), as illustrated in Fig.~\ref{fig:dq-overhead}-(c).
Consequently, fused kernels are effective mainly in memory-bound or small-batch regimes, but become less efficient under compute-bound conditions. 
In large-batch settings, limited on-chip storage capacity (e.g., shared memory) and the compute imbalance often motivate frameworks to split dequantization and GEMM into separate kernels. However, this reintroduces costly HBM write-back and reload operations for intermediate dequantized weights, as shown in Fig.~\ref{kernel}-(a).

Prior studies have sought to reduce dequantization overhead through both hardware~\cite{jang2024figna,fang2025anda} and software~\cite{gptq,park2022lut,jeon2022mr,kim2023squeezellm,lee2024owq,lin2024awq} optimizations. Hardware approaches such as AnDa~\cite{fang2025anda} and FIGNA~\cite{jang2024figna} introduce dedicated FP-INT GEMM units or specialized processing elements to avoid explicit dequantization. Although effective, these designs require intrusive GPU modifications, including new instruction formats and processing-element (PE) designs, limiting scalability and backward compatibility.

In contrast, software approaches preserve hardware transparency, but often rely on specialized kernels. For example, GPTQ~\cite{gptq} uses customized FP-INT GEMM kernels, while LUT-GEMM~\cite{park2022lut} replaces arithmetic operations with lookup-table-based computations. Depending on the method, they may still require non-trivial quantization-dependent kernel tuning or activation-specific LUT generation, which continue to limit their benefits in high-throughput or large-batch cloud inference.

To address these limitations, we propose StreamDQ, a lightweight architectural enhancement for weight-only quantized LLM inference in large-batch, cloud-scale scenarios. As illustrated in Fig.~\ref{kernel}-(b), StreamDQ performs on-the-fly dequantization in the memory subsystem by integrating DeQuantization Blocks (DQBs) into the HBM base die. Each DQB performs dequantization on the memory-read path for standard loads using a lightweight sideband tag while preserving conventional load semantics.
Although StreamDQ follows the general idea of near-memory processing (NMP)~\cite{zhou2023dimm,singh2024dram,ke2021near,alian2018application,jang2019charon}, it avoids key integration overheads of prior GPU-based NMP designs by operating only on weight data and the corresponding $S/Z$ metadata placed within the same pseudo-channel after the GPU’s existing address translation. This avoids runtime data relocation and near-memory VA-to-PA (virtual address-to-physical address) translation overhead. Such integration is becoming increasingly practical because the HBM base die, traditionally fabricated using DRAM process technologies in early generations (e.g., HBM1 and HBM2), has begun adopting more advanced logic technologies in newer designs (e.g., custom HBM for HBM4)~\cite{song2025ai,kim2025anaheim,jun2017hbm}, making it increasingly feasible to integrate lightweight compute logic into the HBM base die within practical power, area, and thermal constraints~\cite{chatterjee2024thermal,glint2024hardware}. In this design, the DQB supports multiple quantized-to-dequantized format conversions (e.g., INT4$\rightarrow$FP16 and INT8$\rightarrow$BF16) through configurable datapaths. It returns dequantized weights through the standard load-response path, allowing tensor-core GEMM to proceed without CUDA-core-based dequantization and improving tensor-core utilization. By offloading dequantization and type conversion to the HBM base die, StreamDQ reduces GPU-side instruction overhead and on-chip traffic. In large-batch regimes with separate dequantization and GEMM kernels, it also avoids extra HBM write-back and reload of intermediate dequantized weights.

StreamDQ can be integrated into existing GPU inference frameworks with modest runtime and GPU-side support, without requiring intrusive changes to the GPU compute datapath. By relocating dequantization to the memory subsystem, StreamDQ tightly couples dequantization with memory access, enabling high-throughput and energy-efficient LLM inference.

This work makes the following contributions:
\begin{itemize}
\item We propose StreamDQ, a lightweight architecture for weight-only quantized LLM inference that performs on-the-fly dequantization in the memory subsystem. It reduces GPU-side dequantization overhead and avoids extra HBM write-back and reload of dequantized weights in large-batch execution with separate dequantization and GEMM kernels.

\item StreamDQ uses a lightweight sideband tag on standard memory read requests and integrates a DQB on each HBM pseudo-channel read path while preserving conventional load semantics, requiring minimal GPU-side changes.

\item StreamDQ supports multiple quantized-to-dequantized format conversions (e.g., INT4$\rightarrow$FP16 and INT8$\rightarrow$BF16) with high area and power efficiency, enabling flexible deployment across diverse LLM configurations.

\item Unlike conventional NMP designs, StreamDQ performs dequantization using only weights and their corresponding $S/Z$ metadata within the same pseudo-channel, avoiding cross-pseudo-channel communication and near-memory VA-to-PA translation overhead while operating within practical HBM base-die power, area, and thermal constraints.

\end{itemize}

%%%%% StreamDQ 이해를 위한 background

\section{Background}

\begin{figure}[t]
\centering
{\includegraphics[width=0.95\columnwidth]{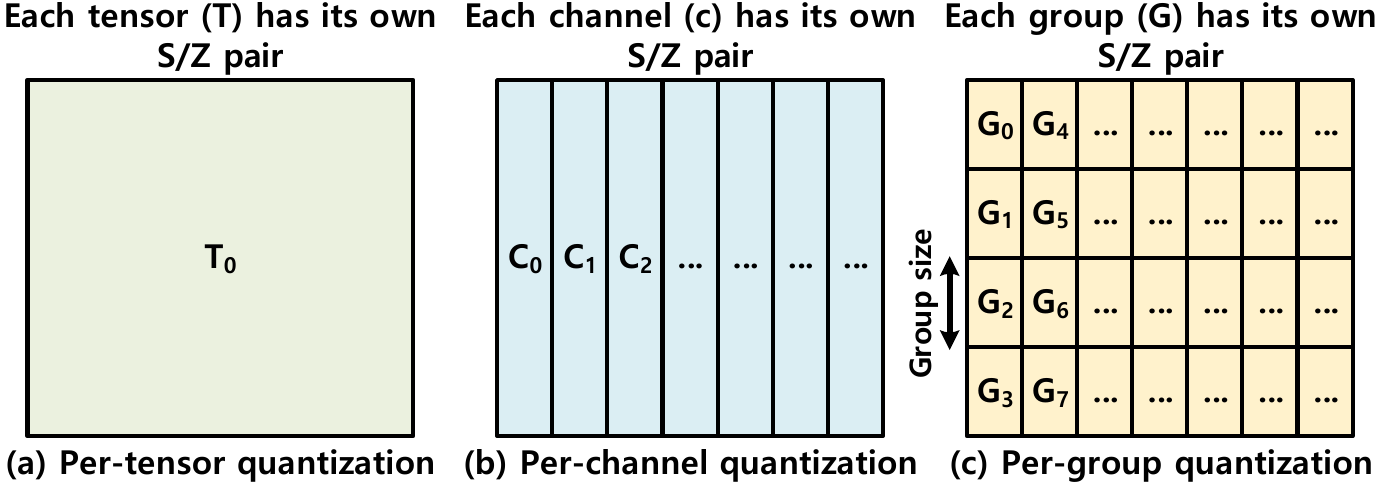}}
\caption{Quantization granularity for weight-only quantization (per-tensor, per-channel, and per-group).}

\label{quantization_granularity}
\vspace{-0.5em}
\end{figure}

\subsection{Quantization Granularity} 
\label{sec:quant_granularity}

Low-bit weight quantization differs not only in bit-width but also in the granularity at which the scaling factor ($S$) and zero-point ($Z$) are defined. This granularity affects both quantization accuracy and how $S/Z$ metadata are accessed during dequantization. Common schemes include per-tensor, per-channel, and per-group quantization, as shown in Fig.~\ref{quantization_granularity}.

\begin{enumerate}
\item \textbf{Per-Tensor Quantization.}
A single $S/Z$ pair is shared by the entire weight tensor (T). This minimizes metadata, but can reduce accuracy at low precision because a single quantization range must represent the entire tensor.

\item \textbf{Per-Channel Quantization.}
One $S/Z$ pair is assigned per output channel (C), improving accuracy at the cost of higher metadata and channel indexing during dequantization.

\item \textbf{Per-Group Quantization.} Each output channel is partitioned into fixed-size groups of $G$ elements, and each group shares one $S/Z$ pair. This better captures local weight-distribution variations and typically improves accuracy at low bit-widths, but requires more $S/Z$ metadata and efficient group-level metadata access.
\end{enumerate}

In low-bit LLM weight-only quantization, per-group quantization is commonly used because it provides a favorable trade-off between accuracy and metadata overhead~\cite{gptq,lin2024awq}. Accordingly, the remainder of this paper focuses on fixed-size per-group quantization and handles $S/Z$ at group granularity.

\begin{figure}[t]
 \centering
 {\includegraphics[width=0.95\columnwidth]{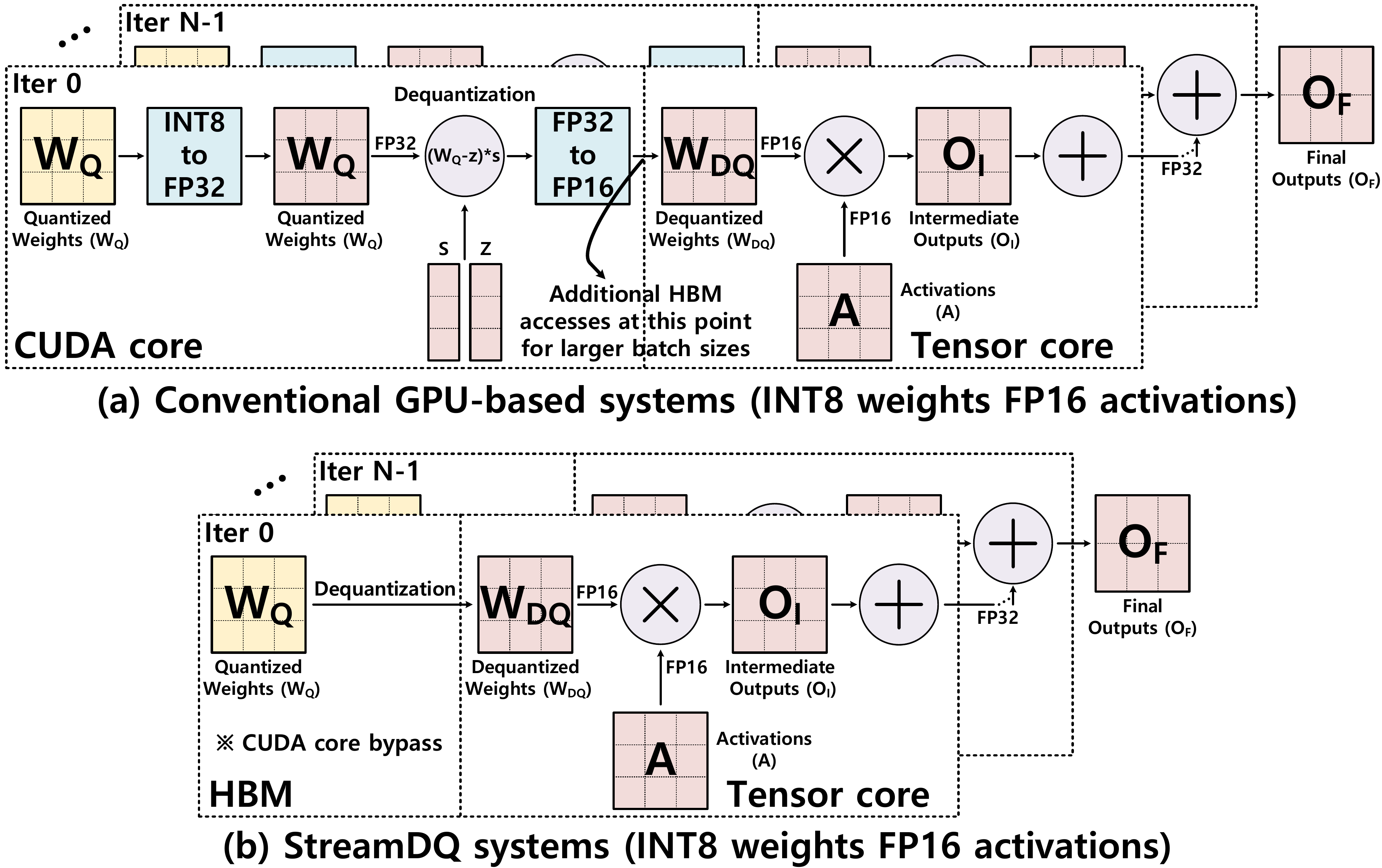}}
 \caption{Comparison of dequantization workflows in (a) conventional GPU-based systems and (b) StreamDQ.}
 \label{dequantprocess}
 \vspace{-1.5em}
\end{figure}

\subsection{Weight-Only Quantized GEMM in Current GPU-based Systems}

In current GPU-based systems, weight-only quantized GEMM requires runtime dequantization of weights before tensor-core GEMM~\cite{lin2024qserve,TensorRT-WoQ}. As illustrated in Fig.~\ref{dequantprocess}-(a), quantized weights are first converted from the stored low-precision format to a higher-precision format, typically FP32 to preserve numerical accuracy, then dequantized using scaling factors and zero-points (Eq.~\ref{equation_quant}), and finally converted to the target compute format (e.g., FP16) before being consumed by tensor cores. Both type-conversion steps, along with the dequantization step, are executed on CUDA cores.

This software-based flow is inefficient because CUDA cores provide much lower throughput and energy efficiency than tensor cores. On NVIDIA A100 GPUs, for example, a CUDA core operation can be up to $50\times$ more costly than a tensor core operation~\cite{lin2024qserve}, leading to pipeline stalls and reduced overall throughput. In large-batch regimes, frameworks may further separate dequantization and GEMM into different kernels to improve tensor-core utilization, but this introduces additional HBM write-back and reload traffic for intermediate dequantized weights, as shown in Fig.~\ref{dequantprocess}-(a).

As shown in Fig.~\ref{dequantprocess}-(b), StreamDQ instead performs on-the-fly dequantization in the memory subsystem. By dequantizing weights inline with memory loads, it bypasses CUDA cores, reduces GPU-side dequantization overhead and on-chip traffic, and eliminates redundant off-chip traffic from intermediate dequantized-weight write-back and reload in large-batch regimes.

\begin{figure}[t]
\centering
{\includegraphics[width=0.95\columnwidth]{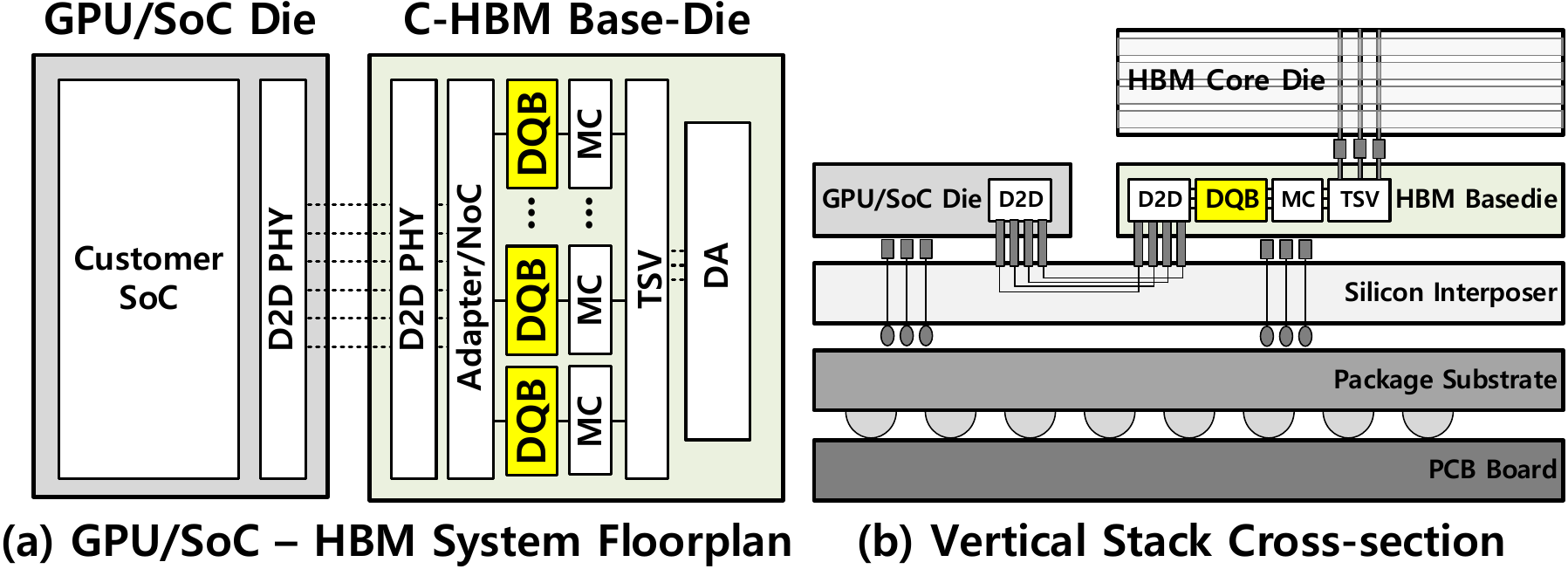}}
\caption{Overall architecture of StreamDQ.}
\label{overallarchitecture}
\vspace{-1.5em}
\end{figure}

%%%%% StreamDQ 아키텍쳐 내용

\section{StreamDQ Architecture}

\subsection{StreamDQ Architecture Overview}

Fig.~\ref{overallarchitecture} illustrates the overall architecture of StreamDQ. StreamDQ is designed as a practical extension to custom HBM (C-HBM) systems~\cite{song2025ai,chatterjee2024thermal}, with the DeQuantization Block (DQB) integrated into the HBM base die rather than in the GPU compute datapath. As shown in Fig.~\ref{overallarchitecture}-(b), each DQB resides within the vertical HBM stack and interfaces with the GPU/SoC die through the standard D2D PHY link, while accesses to the stacked DRAM dies proceed through the memory controller (MC) and TSVs. This organization preserves the conventional HBM access path and remains compatible with established HBM packaging schemes, without disruptive changes to the packaging hierarchy.

To minimize GPU-side modifications, StreamDQ uses a lightweight sideband tag on each memory read request to steer DQB operation without modifying the effective address. This mechanism enables on-the-fly dequantization for standard memory loads, after which the dequantized data is returned through the normal memory response path for tensor-core GEMM execution. Thus, StreamDQ eliminates GPU-side CUDA-core-based dequantization for tagged weight loads while preserving conventional load semantics.

Overall, StreamDQ is designed to operate within the area, power, and thermal constraints of the HBM base die. A DQB is integrated on the read path of each pseudo-channel MC, enabling local on-the-fly dequantization using pseudo-channel-local weights and their corresponding $S/Z$ metadata. Each DQB supports multiple quantized weight formats through configurable conversion logic, providing flexible precision support with modest hardware cost.

\begin{figure}[t]
\centering
\includegraphics[width=0.9\columnwidth]{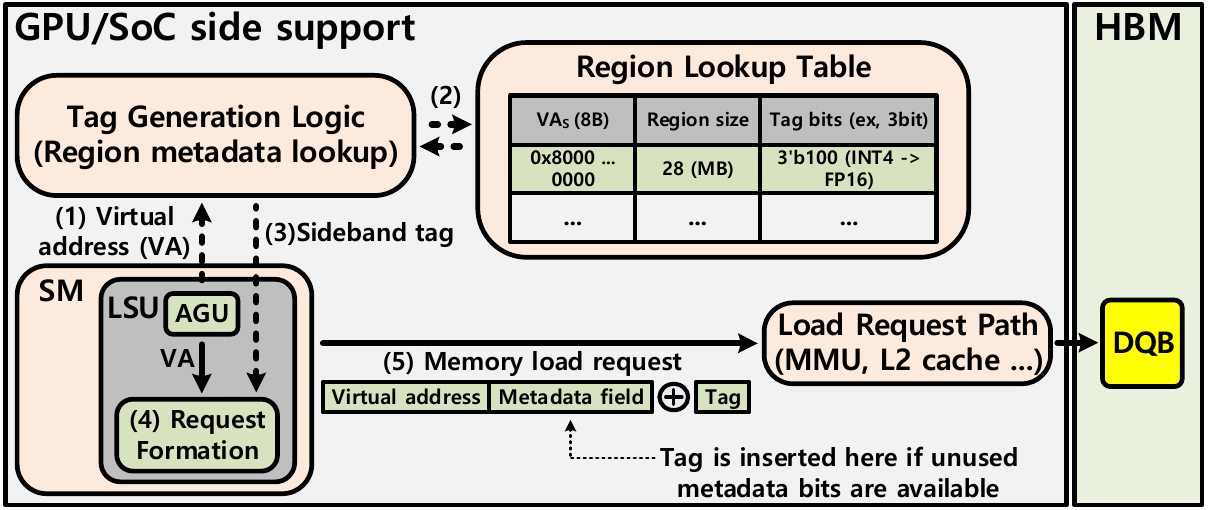}
\caption{
GPU-side support for sideband tag generation using a region-lookup table.
}
\label{gpuside}
\vspace{-0.5em}
\end{figure}

\begin{table}[t]
\centering
\caption{Region-lookup-table size (16\,B per entry), assuming one table entry per quantized linear layer. QL and DB denote quantized linear layers and decoder blocks, respectively.}
\label{tab:region_lookup_table}
\small
\begin{tabular}{lcccc}
\toprule
Model & QLs per DB & DBs & Entries & Lookup-table size \\
\midrule
Llama-3.1-8B~\cite{touvron2023llama} & 7 & 32 & 224 & 3.5\,KB \\
Qwen3-8B~\cite{qwen3}     & 7 & 36 & 252 & 3.9\,KB \\
Mistral-7B~\cite{mistral7b}   & 7 & 32 & 224 & 3.5\,KB \\
\bottomrule
\end{tabular}
\vspace{-1.5em}
\end{table}

\subsection{Sideband Tagging for Dequantization Control}
\label{sec:sideband_encoding}

StreamDQ enables near-memory dequantization by attaching a compact sideband tag to read requests targeting registered quantized-weight regions. The DQB decodes this few-bit tag to select either bypass or a specific conversion mode (e.g., INT4$\rightarrow$FP16). The tag is carried via spare metadata bits in the existing protocol or through a minimal extension to the internal request path. Because the tag is independent of the address, StreamDQ preserves the effective address and requires no ISA-visible changes.

\textbf{Tag Generation and GPU-Side Support.}
As shown in Fig.~\ref{gpuside}, StreamDQ uses a region-lookup table initialized by a privileged runtime layer during model deployment. Each 16\,B entry stores a region start address ($VA_s$), region size, and tag bits.
When an SM issues a load, tag-generation logic checks the virtual address from the address generation unit (AGU) against this table and, upon a match, attaches the corresponding sideband tag to the load request. 
The tag indicates either bypass or a specific conversion mode, while store operations are always issued without tags.
Because this mechanism adds region lookup and propagation of a few tag bits on the request control path, it requires modest GPU-side support and does not alter the standard load path, GPU compute datapath, or conventional load/store semantics.

\textbf{Region Lookup Table Overhead Analysis.}
The region-lookup table requires only 3.5--3.9\,KB for the evaluated models (Table~\ref{tab:region_lookup_table}), assuming one registered quantized-weight region per quantized linear layer. 
This overhead can be further reduced by arranging quantized-weight regions with identical dequantization modes into larger contiguous virtual-address regions at model deployment, thereby reducing the number of table entries. If the number of entries still exceeds hardware capacity, the table can be reprogrammed at layer-group boundaries, potentially using overlap mechanisms (e.g., double-buffering) to reduce reprogramming overhead. Because such updates are infrequent relative to the large number of memory requests issued between them, their overhead can be amortized over end-to-end inference. Even without these optimizations, the required table remains modest relative to on-chip storage resources in modern GPUs.

\begin{figure}[t]
\centering
{\includegraphics[width=\columnwidth]{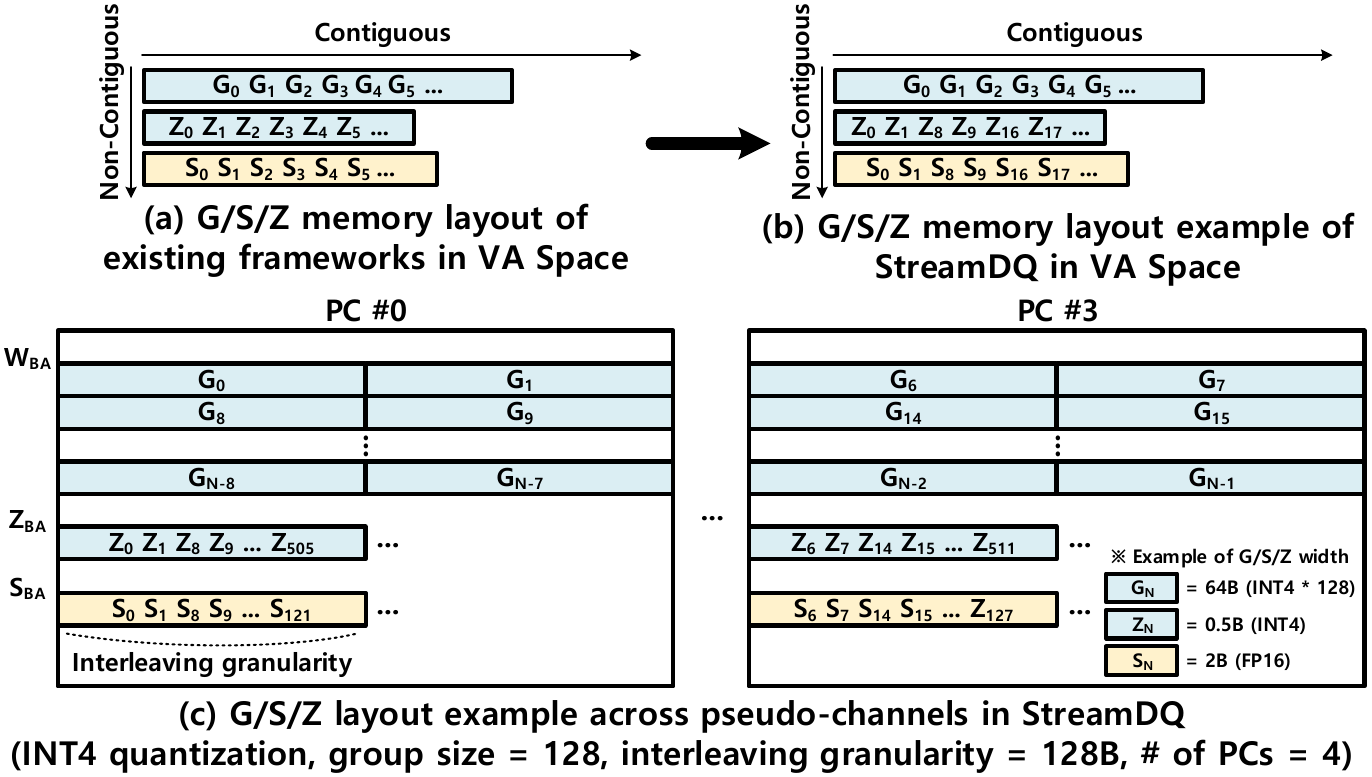}}
\caption{Example of a pseudo-channel-aware layout in StreamDQ, where quantized weight groups and their per-group $S/Z$ metadata are reorganized to reside within the same pseudo-channel for local dequantization.}
\label{layout}
\end{figure}

\begin{table}[t]
\centering
\caption{$S/Z$ replication overhead (\%) relative to the total storage footprint of weights and their corresponding $S/Z$ metadata, across interleaving granularity ($IG$), group size, and quantization bit width, assuming $Z$ matches the weight precision and $S$ is 16-bit.}
\label{tab:sz_replication_overhead}
\begin{tabular}{c|ccc|ccc}
\toprule
\textbf{Quantization} 
& \multicolumn{3}{c|}{\textbf{4-bit}} 
& \multicolumn{3}{c}{\textbf{8-bit}} \\
\midrule
\textbf{IG}
& \textbf{64} & \textbf{128} & \textbf{256}
& \textbf{64} & \textbf{128} & \textbf{256} \\
\midrule
Group size = 64  & 0 & 0 & 0 & 0 & 0 & 0 \\
Group size = 128 & 0 & 0 & 0 & 2.29\% & 0 & 0 \\
Group size = 256 & 1.92\% & 0 & 0 & 3.47\% & 1.16\% & 0 \\
\bottomrule
\end{tabular}
\vspace{-1.5em}
\end{table}

\subsection{Pseudo-Channel-Aware Layout}
\label{sec:layout_mapping}

StreamDQ employs a pseudo-channel-aware layout for quantized weights and their $S/Z$ metadata. The goal is to ensure that each weight group ($G$) and its corresponding metadata are co-located within the same pseudo-channel (PC), allowing the DQB to perform near-memory dequantization without cross-PC communication.

\textbf{Motivation and Transformation.}
Standard quantized layouts typically store weight groups and $S/Z$ metadata in separate contiguous arrays (Fig.~\ref{layout}-(a)). However, modern GPUs commonly interleave memory accesses across HBM pseudo-channels to improve memory-level parallelism. Consequently, a weight group and its $S/Z$ values may map to different PCs, requiring cross-PC accesses for dequantization.

To eliminate this overhead, StreamDQ reorganizes the weight and $S/Z$ layout at model deployment time so that each PC contains both its interleaved weight groups and the corresponding metadata required for local dequantization. For example, if $G_0$, $G_1$, $G_8$, and $G_9$ are mapped to PC \#0, the corresponding $(S,Z)_0$, $(S,Z)_1$, $(S,Z)_8$, and $(S,Z)_9$ are also placed in the same PC, as shown in Fig.~\ref{layout}-(b) and (c). This offline transformation is performed once at deployment and introduces no runtime inference overhead.

\begin{figure}[t]
\centering
{\includegraphics[width=0.92\columnwidth]{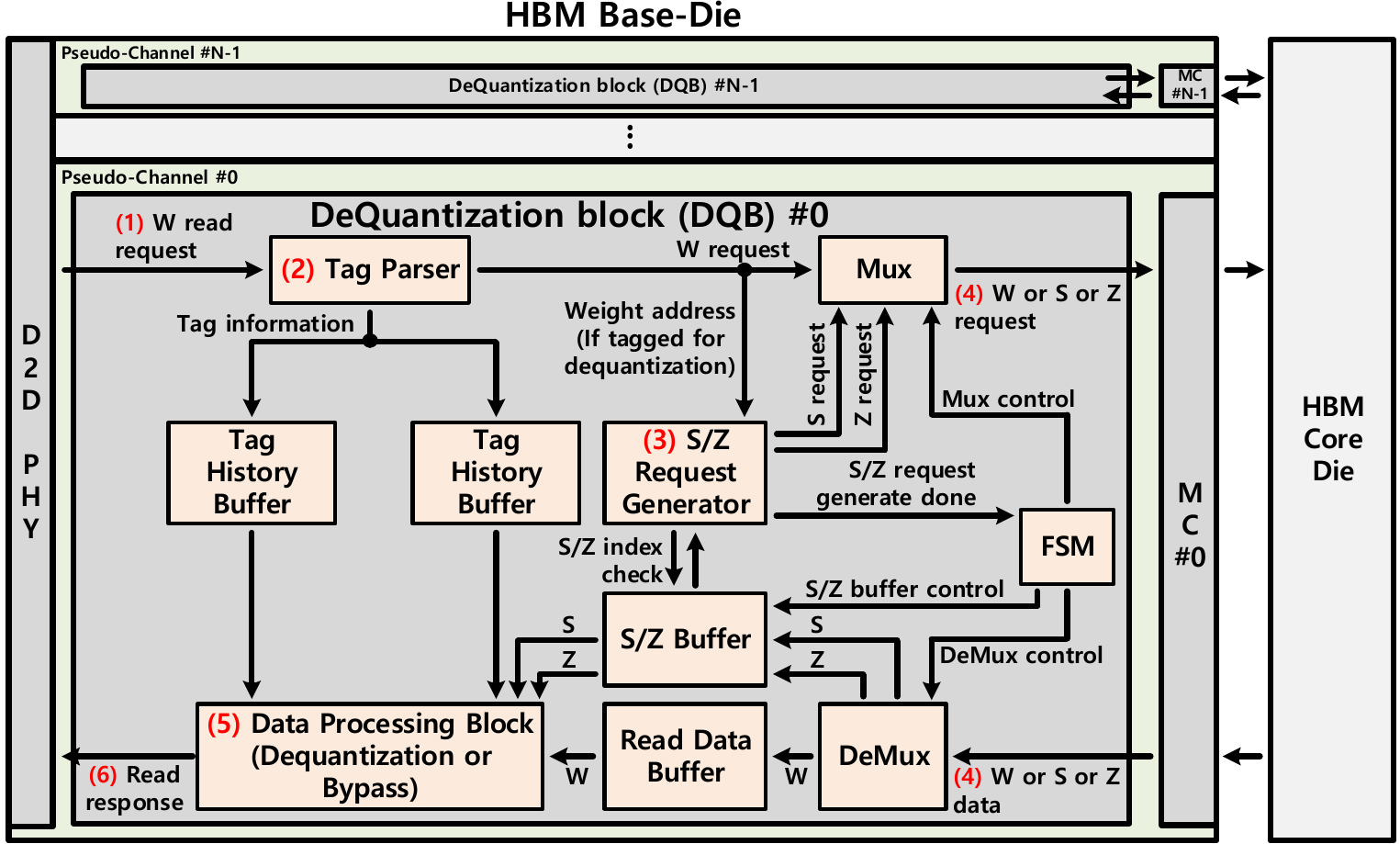}}
\caption{Microarchitecture of a DQB performing on-the-fly dequantization.}
\label{dqb}
\vspace{-1.2em}
\end{figure}

\begin{figure}[t]
\centering
{\includegraphics[width=0.95\columnwidth]{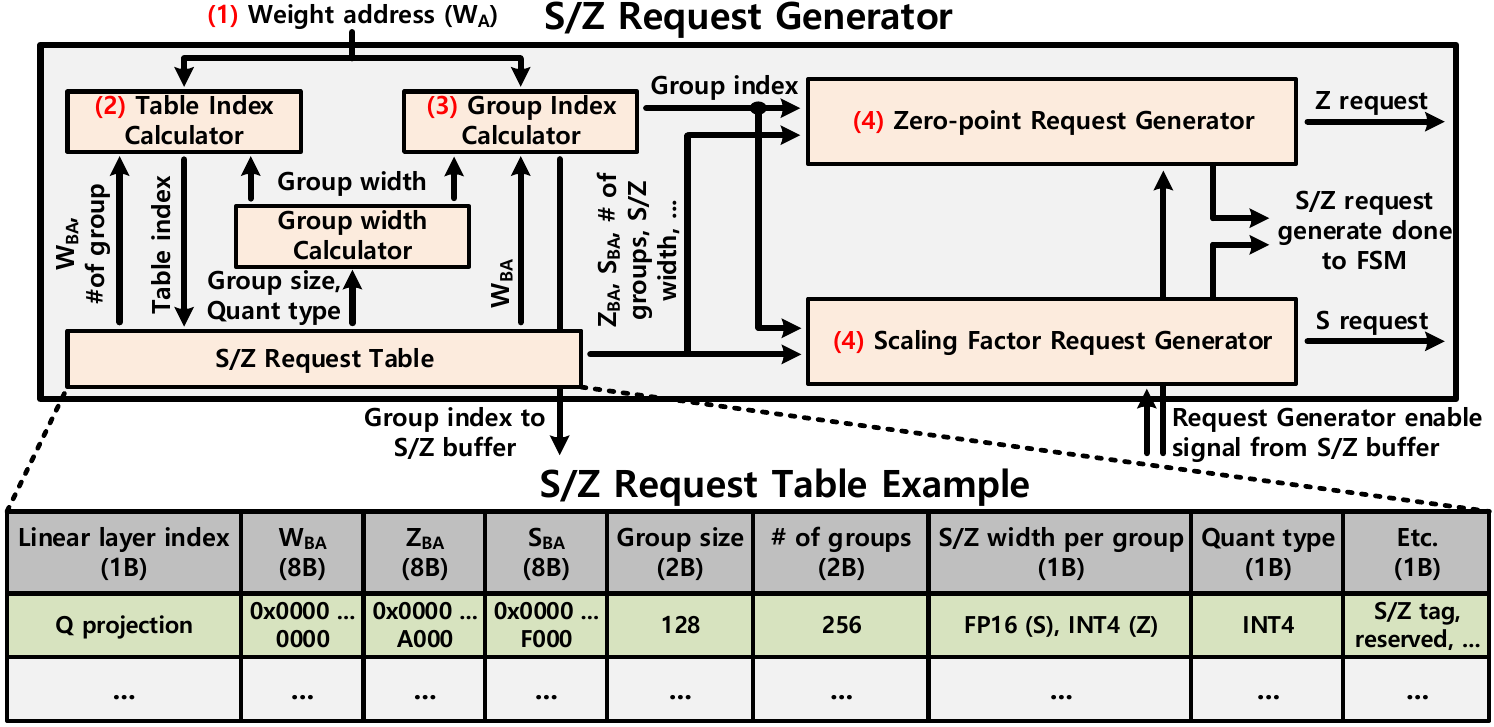}}
\caption{$S/Z$ request generator and an example $S/Z$ request table.}
\label{sz_request}
\vspace{-1.5em}
\end{figure}

\textbf{S/Z Replication Overhead.}
If a group spans multiple PCs (e.g., due to larger $G$ or smaller $IG$), StreamDQ replicates the corresponding $S/Z$ metadata across the involved PCs to preserve local dequantization.
Replication overhead is modest because the metadata footprint is much smaller than that of quantized weights. Table~\ref{tab:sz_replication_overhead} summarizes the replication overhead across quantization bit widths, interleaving granularities ($IG$), and group sizes, relative to the total storage footprint of weights and their corresponding $S/Z$ metadata. Overall, the overhead remains small across practical configurations.

In practice, the design space for $G$ and $IG$ is constrained by accuracy and hardware efficiency. Extremely large group sizes (e.g., $G > 256$) are rarely used in low-bit quantization because they degrade accuracy by forcing one $S/Z$ pair to cover an excessively wide dynamic range. Similarly, $IG$ is typically at least 64\,B in modern GPUs to align with cache-line-sized transfers and maintain burst efficiency; finer granularities increase scheduling complexity and memory stalls. Consequently, for practical LLM deployment scenarios, the overhead of $S/Z$ replication remains negligible.

\subsection{DQB Microarchitecture}
\label{sec:dqb_microarchitecture}

Fig.~\ref{dqb} illustrates the DeQuantization Block (DQB), which consists of three main components: (1) a tag parser for decoding the sideband tag, (2) an $S/Z$ request generator with an on-DQB $S/Z$ buffer, and (3) a data processing block for on-the-fly dequantization. The operation flow of StreamDQ is described in Section~\ref{sec:operation_flow}.

\textbf{Tag Parser and Tag History Buffer.}
For each memory read, the parser decodes the few-bit sideband tag to determine the conversion mode (e.g., INT4$\rightarrow$BF16) or bypass. The effective address is forwarded unchanged to the memory controller. The decoded mode is stored in the tag history buffer, which steers the downstream datapath when the corresponding data returns from HBM.

\begin{table}[t]
\centering
\caption{Fraction of $S/Z$ metadata read requests among total read requests under different group sizes and quantization bit widths, assuming $Z$ matches the weight precision and $S$ is 16-bit.}
\label{tab:request_ratio}
\begin{tabular}{c|cc}
\toprule
\textbf{Group size} & \textbf{4-bit (\%)} & \textbf{8-bit (\%)} \\
\midrule
64  & 7.25 & 4.48 \\
128 & 3.76 & 2.29 \\
256 & 1.92 & 1.16 \\
\bottomrule
\end{tabular}
\vspace{-1.2em}
\end{table}

\textbf{$S/Z$ Request Generator with $S/Z$ Buffer.}
To supply per-group quantization parameters, each DQB includes an $S/Z$ request generator (Fig.~\ref{sz_request}) tightly coupled with a small on-DQB $S/Z$ buffer.
For weight read requests tagged for dequantization, the generator computes the corresponding $S/Z$ addresses using a small per-quantized-linear-layer $S/Z$ request table. 
Each entry encodes the layer parameters needed to derive the corresponding metadata addresses for a given weight address within the same pseudo-channel, including the weight base address ($\mathrm{W_{BA}}$), scaling-factor base address ($\mathrm{S_{BA}}$), zero-point base address ($\mathrm{Z_{BA}}$), and other group configuration parameters. Given a weight address ($\mathrm{W_A}$), the generator first identifies the corresponding layer entry via range matching performed by the table index calculator. The group index is then computed as
\[
\texttt{group\_index} =
\left\lfloor \frac{\mathrm{W_A} - \mathrm{W_{BA}}}{\mathrm{group\_width}} \right\rfloor ,
\]
where $\mathrm{group\_width}$ is the byte stride of one quantization group in the weight array. The $S/Z$ buffer lookup then uses the derived group index within the selected layer entry. On a buffer hit, the cached $S/Z$ metadata are reused. On a miss, the generator derives the metadata addresses using $\mathrm{S_{BA}}$, $\mathrm{Z_{BA}}$, group\_index, and the per-group $S/Z$ widths, and issues standard read requests to the pseudo-channel memory controller to fill the buffer.

Because per-group $S/Z$ metadata is small, each metadata fetch can bring in a contiguous chunk covering many nearby groups; thus, the $S/Z$ buffer can remain small while still achieving a high hit rate. As a result, $S/Z$ reads occur much less frequently than weight reads, as summarized in Table~\ref{tab:request_ratio}. In the large-batch, compute-bound regimes targeted by StreamDQ, this relatively infrequent metadata-fetch overhead is typically amortized and is less likely to become a dominant bottleneck.

A privileged runtime layer initializes the per-layer $S/Z$ request table during model deployment using layout and quantization metadata, including $W/S/Z$ base addresses and group configuration parameters.
Because this setup is performed only once at deployment, it incurs no overhead during runtime inference. Assuming approximately 32\,B per layer entry as shown in Fig.~\ref{sz_request}, the evaluated models (Table~\ref{tab:region_lookup_table}) require only about 7--8\,KB of storage per DQB. This modest footprint leaves room for moderate capacity scaling while keeping the per-DQB overhead low. If the provisioned capacity is still insufficient, the table can be reloaded at coarse-grained layer-group boundaries, potentially with overlap mechanisms (e.g., double buffering), and the associated overhead can be amortized over many subsequent weight accesses.

\begin{figure}[t]
\centering
{\includegraphics[width=0.95\columnwidth]{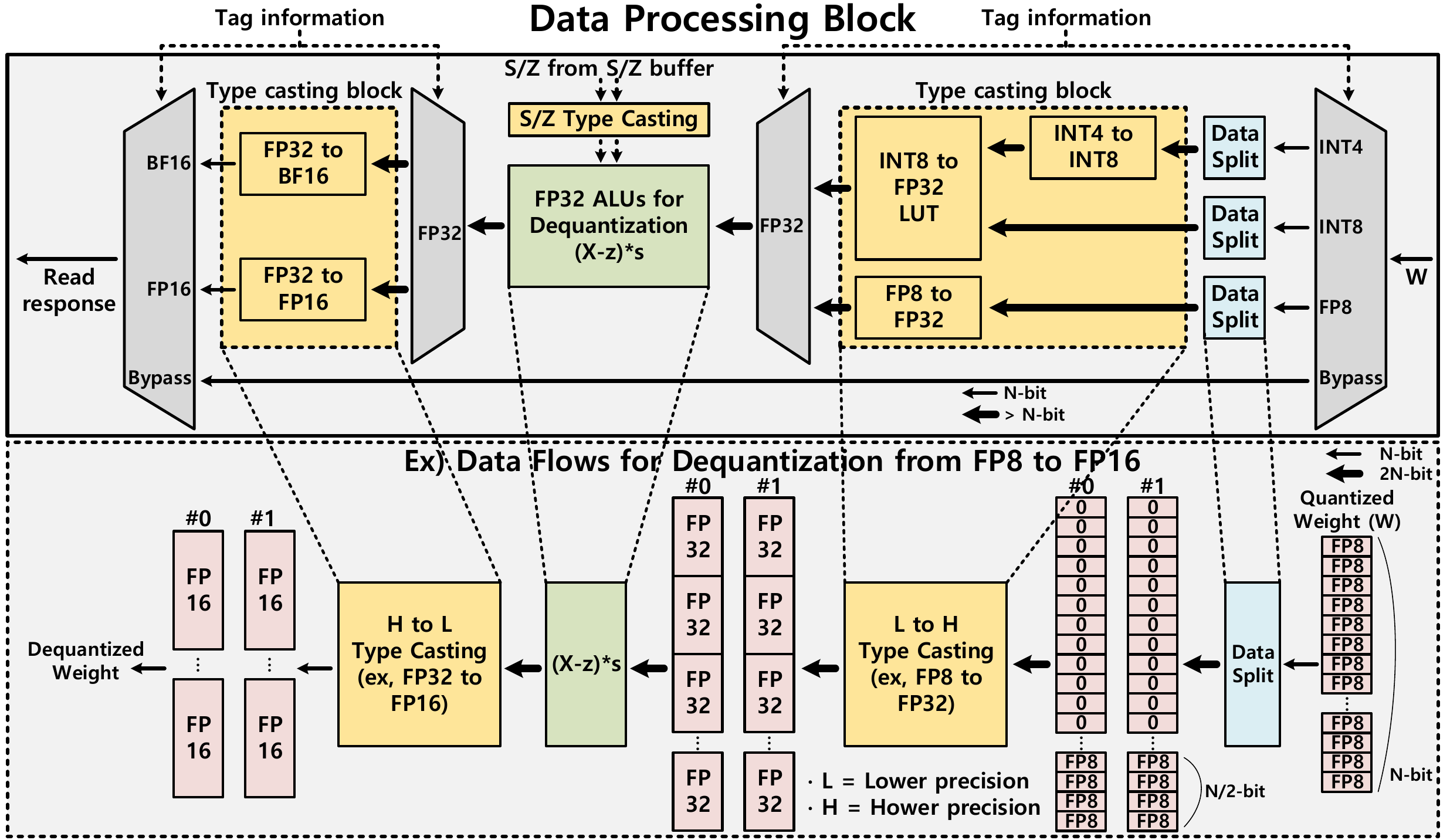}}
\caption{DQB data processing block for on-the-fly type conversion and dequantization, with a bypass path and an example FP8-to-FP16 data flow.}
\label{dataprocessingblock}
\vspace{-1.5em}
\end{figure}

\textbf{Data Processing Block.}
Fig.~\ref{dataprocessingblock} shows the structure of the data processing block and an example FP8$\rightarrow$FP16 dequantization data flow. Based on the stored tag, the block either forwards data through a bypass path or performs type conversion and dequantization using buffered $S/Z$ values. The dequantization arithmetic reuses shared FP32 ALUs across the supported input formats, while lightweight type-conversion paths handle format-specific conversion. To handle bit-width expansion (e.g., 4-bit to 16-bit), it employs a lightweight data-splitting stage and an internal datapath wider than the external interface. Simple mux/demux logic, controlled by the tag information, routes data through the appropriate bypass or dequantization path.

\subsection{StreamDQ Operation Flow} 
\label{sec:operation_flow} 
The operation flow of a StreamDQ-enabled read transaction is illustrated in Fig.~\ref{dqb} and proceeds as follows:
\begin{enumerate} 
\item \textbf{Read Request.} The GPU issues a standard load. If the address matches a registered quantized-weight region, the GPU attaches a few-bit sideband tag to the read request, which is delivered to the DQB via the D2D PHY.
\item \textbf{Tag Parsing.} Upon receiving a tagged read request, the parser decodes the sideband tag to determine the conversion mode or bypass. The decoded tag information is stored in the tag history buffer to steer the datapath when the read data returns from memory.
\item \textbf{$S/Z$ Request Generation.} For per-group dequantization, the DQB derives the group index from the weight address and checks the local $S/Z$ buffer. On a miss, the $S/Z$ request generator issues additional HBM read requests to fetch the required $S/Z$ values.

\item \textbf{Memory Access and Data Arrival.} The DQB forwards the weight request and any required $S/Z$ requests to the memory controller. A lightweight FSM arbitrates these requests to ensure that the required $S/Z$ metadata is available before the associated weight data is processed.

\item \textbf{On-the-fly Dequantization.} On the read path, the DQB either performs dequantization using buffered $S/Z$ values or forwards the data through the bypass path, depending on the stored tag.

\item \textbf{Read Response.} The processed data returns to the GPU via the D2D PHY through the standard load response path, preserving conventional load semantics.

\end{enumerate} 
By offloading dequantization to the memory subsystem while preserving standard load semantics, StreamDQ reduces GPU-side instruction overhead and register pressure, improving tensor-core utilization during quantized LLM inference.

\begin{figure}[t]
\centering
{\includegraphics[width=0.93\columnwidth]{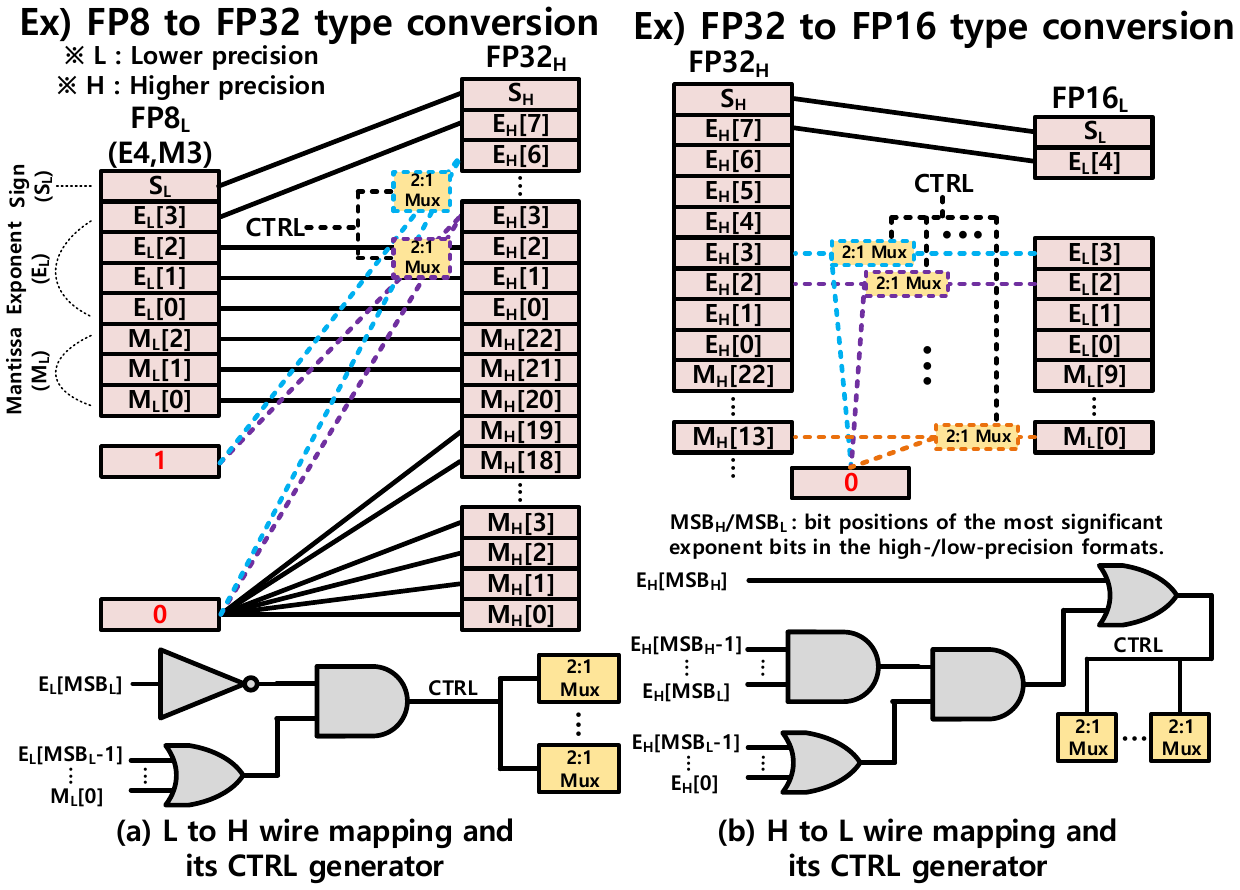}}
\caption{Wire-mapping-based FP-to-FP type-conversion mechanism for normalized values and zero.}
\label{fptofp}
\vspace{-1.2em}
\end{figure}

\subsection{Efficient Support for Diverse Formats}
StreamDQ supports multiple quantized-to-dequantized format pairs while staying within the tight area and power budgets of the HBM base die. Rather than instantiating separate dequantization datapaths for each format pair, the DQB reuses shared FP32 dequantization ALUs across input formats (Fig.~\ref{dataprocessingblock}). Type conversion is implemented using lightweight hardware mechanisms, enabling efficient on-the-fly dequantization with low logic overhead.

\textbf{FP-to-FP Type Conversion.}
For floating-point format conversion (e.g., FP8$\rightarrow$BF16), StreamDQ uses a wire-mapping datapath for normalized values and zero, avoiding ALUs or shifters for exponent bias adjustment~\cite{ootomo2023custom,saldanha2009float}. As shown in Fig.~\ref{fptofp}, for L$\rightarrow$H conversion, it maps the sign bit and the most significant exponent bit (E[MSB]) of $FP_L$ to the corresponding positions in $FP_H$, places the remaining exponent bits of $FP_L$ in the lower exponent positions of $FP_H$ from the least significant bit, and places the mantissa bits of $FP_L$ in the most significant mantissa positions of $FP_H$. Lightweight 2:1 multiplexers controlled by CTRL fill the unmatched higher exponent bits of $FP_H$ with either `0' or `1', while the remaining lower mantissa bits are zero-filled. CTRL is `0' when the input is zero or when the most significant exponent bit of $FP_L$ is 1, and `1' otherwise.

For H$\rightarrow$L conversion, StreamDQ uses the reverse wire-mapping flow. The sign bit and E[MSB] are directly mapped to their corresponding positions in $FP_L$, while the remaining mapped bits pass through a small set of 2:1 multiplexers controlled by CTRL, which select either the corresponding wire-mapped bit or zero. With round-toward-zero and flush-to-zero for denormalized results, CTRL forwards the wire-mapped bit only when the type-conversion result is a normalized value, and selects zero otherwise.

By replacing ALU- or shifter-based conversion with wire mapping and lightweight multiplexers, StreamDQ provides an area- and power-efficient FP-to-FP conversion mechanism suitable for the tight integration constraints of the HBM base die.

\begin{figure}[t]
\centering
{\includegraphics[width=0.95\columnwidth]{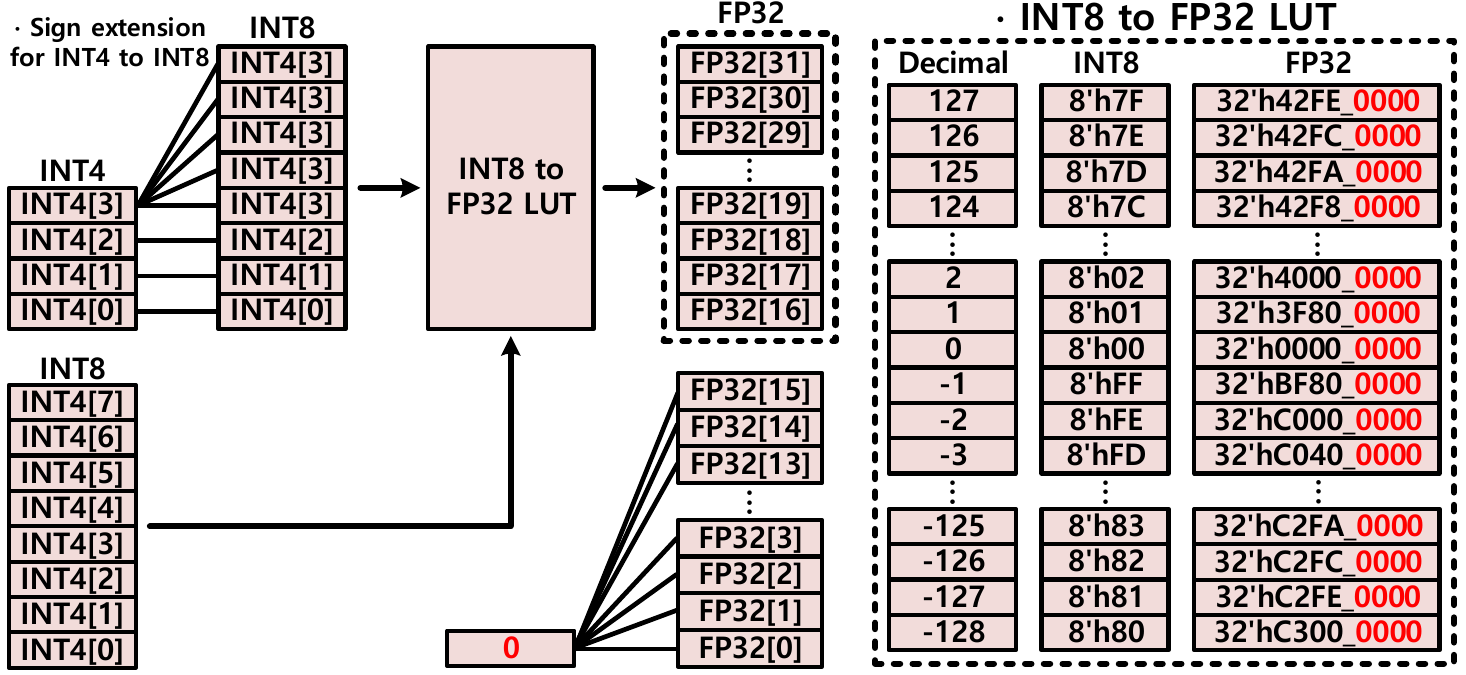}}
\caption{INT-to-FP type conversion using a shared LUT and zero-padding optimization.}
\label{inttofp}
\vspace{-1.5em}
\end{figure}

\textbf{INT-to-FP Type Conversion.}
For integer inputs, StreamDQ uses a shared LUT-based INT$\rightarrow$FP conversion structure with zero-padding optimization. 
Instead of maintaining separate LUTs for each supported integer bit-width, it builds a single LUT for the largest supported signed format (e.g., INT8) and reuses it for smaller signed formats (e.g., INT4) via sign-bit extension into the added upper bits, as illustrated in Fig.~\ref{inttofp}.
To reduce LUT cost further, StreamDQ exploits the fact that INT8-to-FP32 conversion has no fractional part. The lower 16 bits of the FP32 output are hardwired to zero, so the LUT stores only the upper 16 bits. Thus, the 8-bit-to-32-bit lookup is effectively reduced to an 8-bit-to-16-bit LUT. This preserves exact numerical accuracy while reducing LUT storage and energy.

\subsection{Interleaving-Aware DQB Deployment and Scalability}

\begin{figure}[t]
\centering
{\includegraphics[width=0.9\columnwidth]{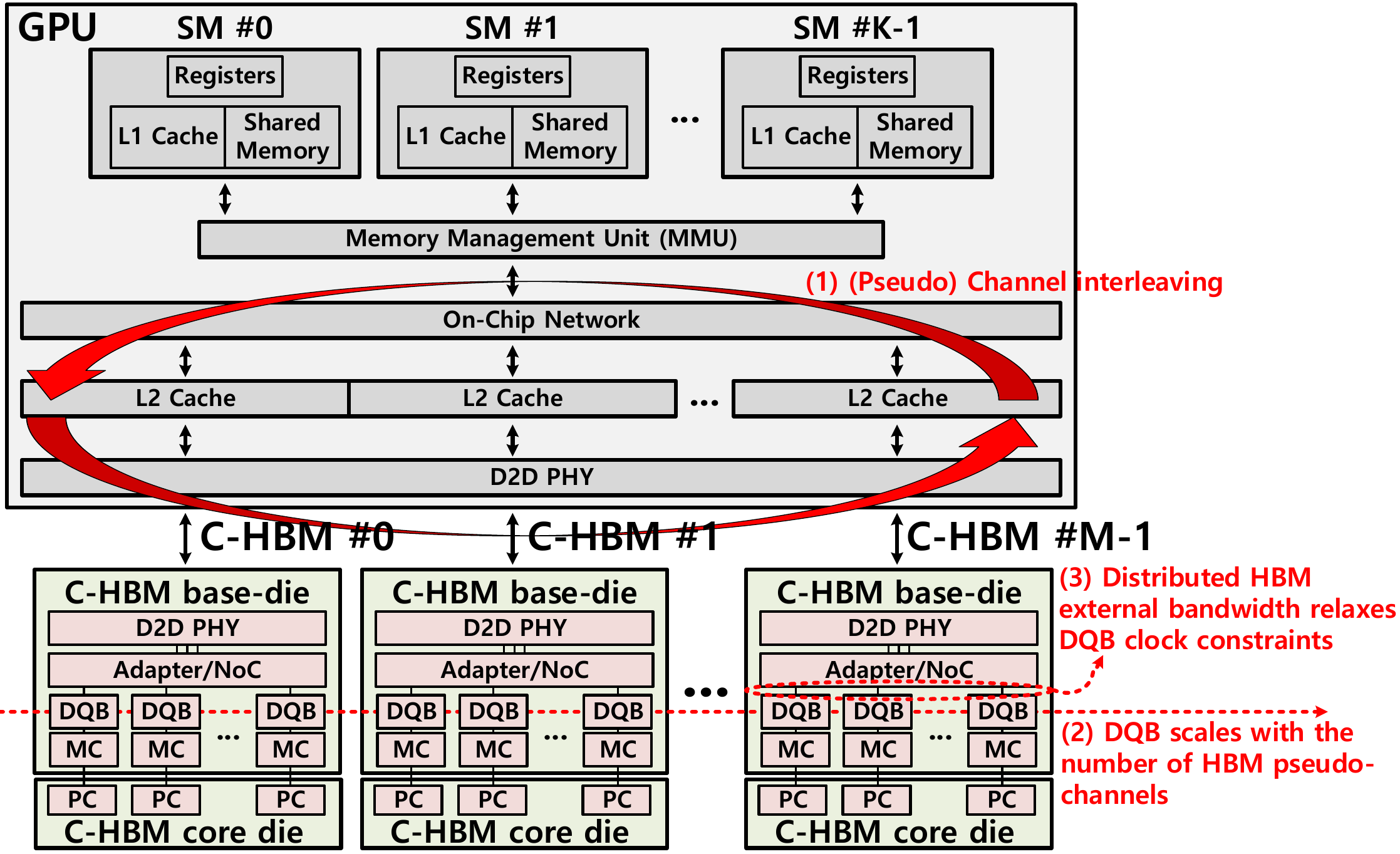}}
\caption{Interleaving-aware deployment of DQBs across HBM pseudo-channels (PCs), with one DQB on the read path of each pseudo-channel memory controller (MC)}
\label{placement}
\vspace{-0.5em}
\end{figure}

\textbf{Interleaving-Aware Deployment.}
To align with pseudo-channel-level interleaving commonly used in modern GPU-HBM systems, StreamDQ deploys one DQB on the read path of each pseudo-channel memory controller (MC), as illustrated in Fig.~\ref{placement}. Although the exact interleaving policies of modern GPUs are proprietary, pseudo-channel interleaving is widely used to distribute memory traffic and improve memory-level parallelism. This organization enables parallel dequantization along the memory access path while keeping each DQB local to its assigned pseudo-channel.

\textbf{Scalability.}
As HBM generations evolve, both bandwidth and the number of pseudo-channels vary across designs. Because each DQB operates independently within its local pseudo-channel, the aggregate dequantization throughput scales with the number of pseudo-channels. Moreover, since HBM traffic is distributed across pseudo-channels, each DQB needs to sustain only the bandwidth of its local pseudo-channel and can therefore operate at a relatively low clock frequency. This relaxes timing constraints and allows StreamDQ to scale with future HBM bandwidth through modest local scaling rather than major architectural changes.

Supporting future low-precision formats may require format-specific hardware logic. However, some existing components, such as the shared FP32-ALU-based dequantization arithmetic, can still be reused, with only lightweight extensions to the few-bit tag encoding and format-specific control, or even without increasing tag width if unused encodings remain. Therefore, the overhead of extending StreamDQ to additional low-precision formats is expected to remain modest.

\begin{figure}[t]
\centering
{\includegraphics[width=\columnwidth]{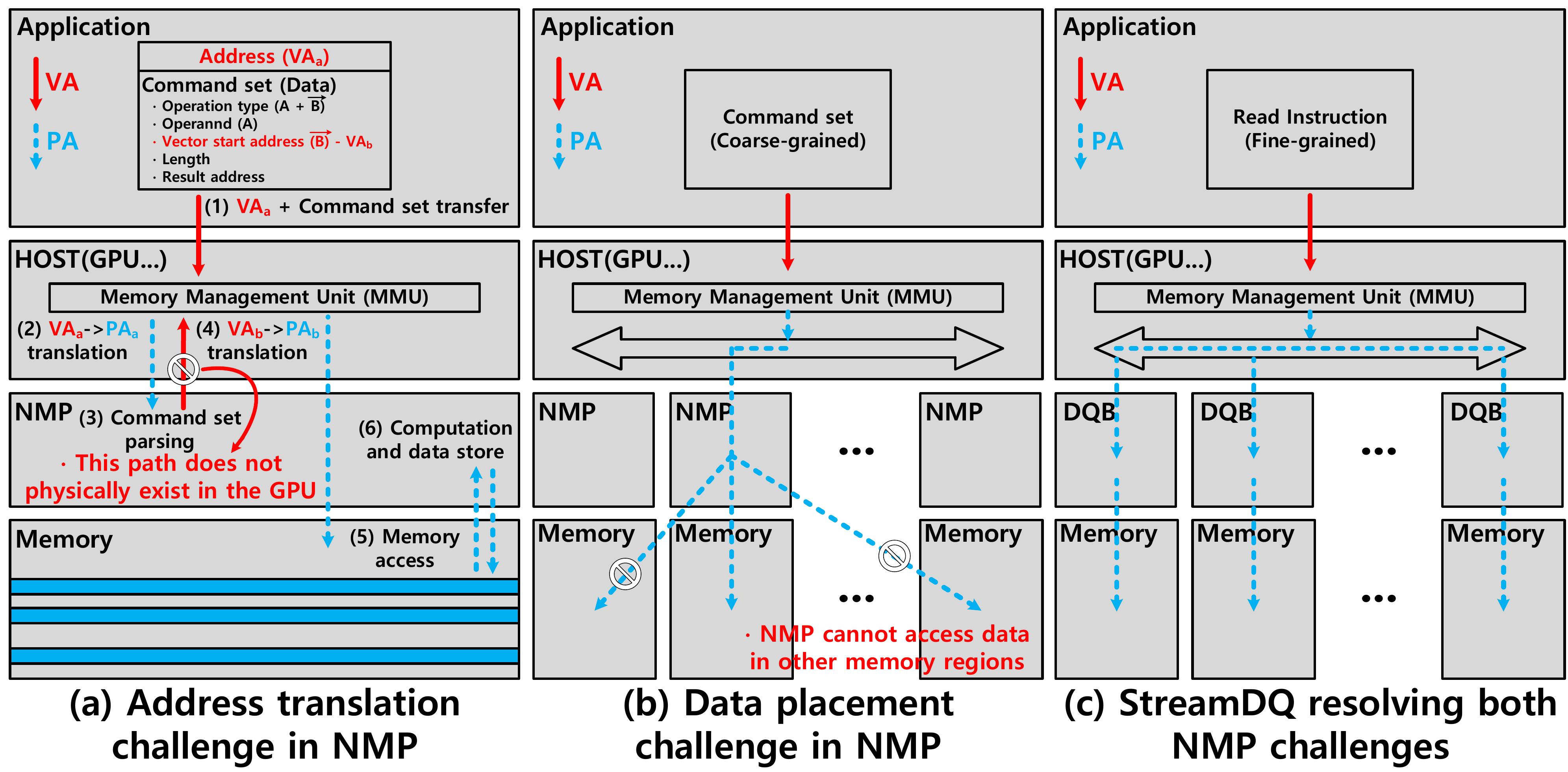}}
\caption{Address translation and data placement challenges in GPU-based NMP, and StreamDQ’s approach.}
\label{nmp}
\vspace{-1.5em}
\end{figure}

\section{Overcoming NMP Challenges in StreamDQ}
In GPU systems, enabling near-memory dequantization requires addressing two fundamental NMP challenges: address translation~\cite{ke2020recnmp,picorel2017near} and data placement~\cite{li2024stream,kim2017toward}.

\textbf{Address Translation Challenge.}
Conventional NMP is difficult to integrate into GPU systems because near-memory logic operates in the physical-address (PA) domain, whereas GPU requests originate in the virtual-address (VA) space. As shown in Fig.~\ref{nmp}-(a), the command-set address ($VA_a$) is translated by the GPU MMU, but an embedded operand address ($VA_b$) bypasses the MMU because it is carried as data, requiring translation support in the near-memory unit. Exposing the GPU MMU to near-memory logic would require substantial MMU-interface and control-path integration, while local translation support would incur substantial area and power cost.

\textbf{Data Placement Challenge.}
GPUs interleave data across channels and stacks to maximize throughput, whereas each NMP unit is local to a specific memory region. As shown in Fig.~\ref{nmp}-(b), the operands required for a computation may therefore span multiple regions, preventing a single NMP unit from accessing the full operand set locally. This mismatch would require additional cross-region communication support, including interconnect and memory-controller mechanisms to fetch operands from multiple channels or stacks.

\textbf{StreamDQ’s Approach.}
As shown in Fig.~\ref{nmp}-(c), StreamDQ addresses both challenges by leveraging the GPU’s standard MMU path and confining dequantization to pseudo-channel-local weights and metadata. Each DQB is tightly coupled to a pseudo-channel, uses physical addresses after MMU translation, and accesses only local data. This eliminates additional near-memory address translation and avoids cross-channel communication, yielding lower integration complexity than conventional GPU-based NMP.

%%%%% 실험 방법

\section{Evaluation Setup}

We evaluate StreamDQ from two perspectives against state-of-the-art software-based dequantization approaches: (1) mixed-precision GEMM (mpGEMM) performance and energy consumption, including both dequantization and GEMM costs, and (2) end-to-end LLM inference performance. For performance and power analysis, we use StreamDQ-Sim, an in-house simulator for StreamDQ.

\subsection{Benchmarks}

All runtime evaluations (Sections~\ref{sec:mixed_precision_gemm} and~\ref{sec:eval_inf}) were conducted on an NVIDIA A100 PCIe 40\,GB GPU with five HBM2 stacks. The software stack consists of vLLM~0.11.0 (v1 engine)~\cite{vllm-github}, PyTorch~2.8.0, and CUDA~12.8. We evaluate three representative LLMs (LLaMA-3.1-8B-Instruct~\cite{touvron2023llama}, Qwen3-8B~\cite{qwen3}, and Mistral-7B-Instruct-v0.3~\cite{mistral7b}) and their quantized variants (W4A16 and W8A16).

For mpGEMM evaluation, we compare StreamDQ against GPTQ, AWQ-v1, AWQ-v2, and TorchAO. GPTQ and AWQ-v1 are integrated into vLLM~\cite{vllm-github}, AWQ-v2 is the latest fused kernel released by the AWQ authors~\cite{awq-github}, and TorchAO is evaluated using its public implementation~\cite{torchao}. We evaluate four quantized-to-dequantized format pairs: INT4$\rightarrow$FP16, INT4$\rightarrow$BF16, INT8$\rightarrow$FP16, and INT8$\rightarrow$BF16.

For end-to-end evaluation, we compare StreamDQ with AWQ~\cite{lin2024awq} and GPTQ~\cite{gptq} across a range of batch sizes using inference latency and decode throughput.

\subsection{Simulation Methodology}
We developed StreamDQ-Sim, an in-house simulator for evaluating StreamDQ performance and energy when integrated into the HBM base die. StreamDQ-Sim consists of three components: (1) trace extraction and conversion, (2) a modified Accel-Sim~\cite{accelsim} incorporating StreamDQ’s near-memory dequantization behavior (StreamDQ-enabled Accel-Sim), and (3) an analysis module calibrated to real GPU measurements.

Execution traces are collected using the NVBit-based Accel-Sim tracer~\cite{nvbit} from full-precision GEMM (fpGEMM) kernels (e.g., FP16), then converted into mpGEMM workloads (e.g., W4A16) and simulated with StreamDQ-enabled Accel-Sim.

%In Accel-Sim, each LDG (load from global) read is modeled as N fixed-granularity memory transactions, and the corresponding DRAM return is modeled as N response transactions. When StreamDQ is incorporated into Accel-Sim, the GPU issues read requests for quantized data, resulting in fewer request transactions, while near-memory dequantization in HBM expands the data and returns N response transactions in the dequantized format. Accel-Sim processes these request/response messages using queue-based structures.

Performance and power are modeled with Accel-Sim and AccelWattch~\cite{accelwattch}, respectively. To estimate real-GPU behavior, StreamDQ-Sim calibrates simulated mpGEMM results using the relationship between simulated fpGEMM results from Accel-Sim and measured fpGEMM performance and power on an NVIDIA A100 collected with Nsight Compute~\cite{nsight-compute} and NVML~\cite{nvml}. As demonstrated in prior work~\cite{accelsim,accelwattch}, the strong correlation between simulated and measured fpGEMM results serves as a scaling factor for projecting StreamDQ’s real-GPU performance and power.

Because AccelWattch supported power modeling only up to V100, we extended it with A100-specific architectural and power parameters. The resulting model shows strong agreement with real-silicon measurements: for FP16 and BF16 GEMM, the Pearson Correlation Coefficients (PCCs) are 0.833 and 0.866, and the Mean Absolute Percentage Errors (MAPEs) are 28.67\% and 27.26\%, respectively. For larger GEMM configurations ($M \leq 64$, $K=N=4096$), the MAPEs decrease to 13.35\% and 7.67\%, respectively.

Beyond kernel-level evaluation, StreamDQ-Sim also estimates end-to-end inference gains. Using Nsight Systems~\cite{nsight-systems}, we profile the mpGEMM fraction in representative LLM traces and apply the simulated mpGEMM speedups to estimate system-level improvements from StreamDQ.

\begin{table}[t]
\centering
\caption{Public HBM3 stack parameters~\cite{park2022192,NVIDIAH100} used as a reference point for the area/power feasibility of integrating DQBs into the HBM base die.}
\label{table1}
\begin{tabular}{ll}
\toprule
\textbf{Parameter} & \textbf{Value} \\
\midrule
HBM base-die dimension & 11\,mm $\times$ 11\,mm = 121\,mm$^2$ \\
Effective HBM frequency & 2619\,MHz \\
Number of pins & 1024 \\
Number of channels & 16 \\
Pseudo-channels per channel & 2 \\
\bottomrule
\end{tabular}
\end{table}

%%%%% 실험결과
\section{Evaluation Results}

\subsection{Area and Power Overhead of the DQB IP}

To evaluate the feasibility of integrating StreamDQ into modern HBM systems, we performed pre-CTS (clock-tree synthesis) RTL synthesis of the DQB using a 12\,nm CMOS logic process and Synopsys Design Compiler~\cite{synopsys}. Because detailed public HBM4 stack specifications are not yet fully available, we use publicly available HBM3 parameters (Table~\ref{table1}) as a reference for stack-level area and power overhead estimation. Table~\ref{table2} summarizes the key DQB design parameters.

The DQB uses a 3-bit sideband tag for dequantization control. Each DQB provides a 512-bit external interface per pseudo-channel and a 1024-bit internal datapath, operating at 328\,MHz to sustain high-throughput interleaved HBM accesses.

A single DQB occupies 0.127\,mm$^2$. Under the 32-pseudo-channel-per-stack configuration of the reference HBM3 stack, the total DQB area corresponds to approximately 3.36\% of the base-die area. Power analysis at a 20\% input toggle rate reports 0.355\,W per DQB, corresponding to 11.36\,W for an entire stack. The relative area and power overheads of StreamDQ are expected to decrease further as HBM base-die implementations adopt more advanced logic processes, and to decrease further still if the $S/Z$ tables and buffers are implemented in SRAM rather than logic-based buffers.

\begin{table}[t]
\centering
\caption{Design parameters of a DQB}
\label{table2}
\begin{tabular}{ll}
\toprule
\textbf{Parameter} & \textbf{Value} \\
\midrule
Technology & 12\,nm CMOS logic process \\
Supported conversion & \{INT4, INT8, FP8\} $\rightarrow$ \{FP16, BF16\} \\
Tag bit width & 3 bits \\
Tag bit information & 3'b000 = Bypass \\
& 3'b001 = INT4 $\rightarrow$ BF16 \\
& 3'b010 = INT8 $\rightarrow$ BF16 \\
& 3'b011 = FP8 $\rightarrow$ BF16 \\
& 3'b100 = INT4 $\rightarrow$ FP16 \\
& 3'b101 = INT8 $\rightarrow$ FP16 \\
& 3'b110 = FP8 $\rightarrow$ FP16 \\
& 3'b111 = Reserved for future use \\
DQB interface width & 512 bits \\
Internal datapath width & 1024 bits \\
Operating clock frequency & 328\,MHz \\
S/Z request table size & 8\,KB \\
S/Z buffer size & 4\,KB \\
Area & 0.127\,mm$^2$ \\
Power & 0.355\,W (at 20\% input toggle rate) \\
\bottomrule
\end{tabular}

\end{table}

\subsection{Thermal Feasibility of StreamDQ in the HBM Base Die}

Thermal characterization of the HBM base die is important because excessive thermal coupling to the DRAM layers can cause data-retention failures (e.g., bit flips)~\cite{cho2006analysis,weis2015retention}. Therefore, only low-power IP blocks with limited thermal impact are suitable for base-die integration.
We evaluated the thermal feasibility of StreamDQ using FloTHERM~\cite{flotherm}. While the DQB power values are derived from our 12\,nm RTL synthesis (Table~\ref{table2}), the thermal analysis uses a representative custom-HBM-like logic-base-die model. To reflect the more advanced logic technology expected for such base dies, DQB power density is scaled under a proportional power-scaling assumption. The model includes power maps for the D2D PHY, memory controllers (MCs), NoC, and DQBs (Fig.~\ref{overallarchitecture}).

\begin{figure}
    \centering
    \includegraphics[width=0.95\linewidth]{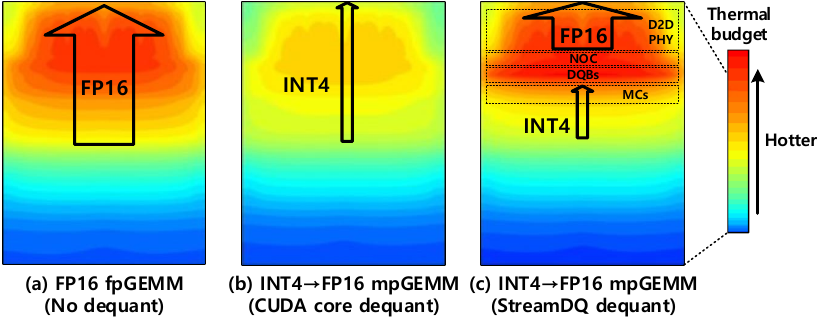}
\caption{HBM base-die thermal distributions for (a) FP16 fpGEMM without dequantization, (b) INT4-FP16 mpGEMM with CUDA-core dequantization, and (c) INT4-FP16 mpGEMM with StreamDQ dequantization.}
    \label{thermal}
    \vspace{-1.2em}
\end{figure}

Fig.~\ref{thermal} compares three scenarios:
\begin{enumerate}
\item \textbf{GPU-FP16 (no dequantization).} FP16 weights are fetched and processed on the GPU. The resulting base-die temperature profile remains within the thermal budget of the custom HBM stack, leaving sufficient DRAM retention margin.
\item \textbf{GPU-W4 (CUDA dequantization).} INT4 weights are dequantized on CUDA cores. Because less data is transferred from HBM to the GPU than in Fig.~\ref{thermal}-(a), the resulting thermal profile is slightly lower.
%INT4 weights are dequantized on CUDA cores. Due to CUDA-core bottlenecks and reduced data size, effective bandwidth is lower than in Figure~\ref{thermal}-(a), producing a slightly reduced thermal profile.
\item \textbf{HBM-W4 (StreamDQ).} Dequantization is performed in the base die. The FP16 ``upper'' path exhibits a thermal profile similar to Fig.~\ref{thermal}-(a), while the INT4 ``lower'' path follows Fig.~\ref{thermal}-(b). Although the DQB region shows a small localized temperature rise, it remains well below the base-die thermal limit.
\end{enumerate}

Overall, StreamDQ introduces only minor localized heating and remains within the thermal constraints of the HBM base die. These results indicate that StreamDQ can be integrated into the HBM base die without significantly reducing the DRAM reliability margin.

\begin{figure*}[t]
    \centering
    \begin{subfigure}[t]{0.250\textwidth}
        \centering
        \includegraphics[width=1.0\linewidth]{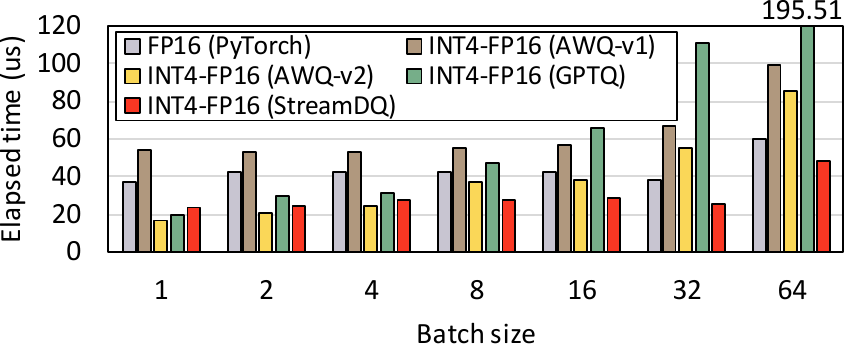}
        \caption{INT4-FP16}
        \label{fig:perf_int4-fp16}
    \end{subfigure}
    ~ 
    \begin{subfigure}[t]{0.250\textwidth}
        \centering
        \includegraphics[width=1.0\linewidth]{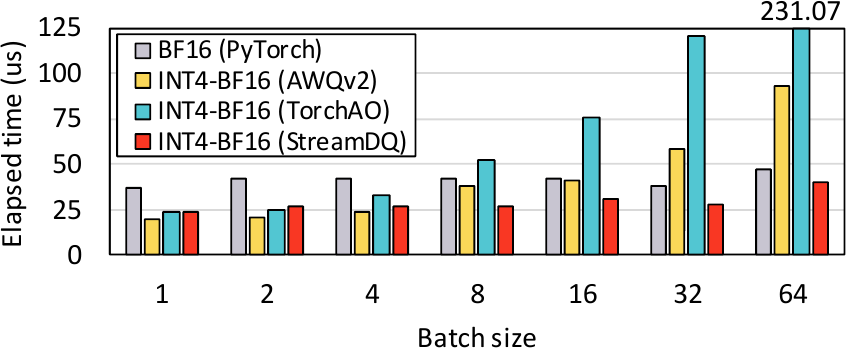}
        \caption{INT4-BF16}
        \label{fig:perf_int4-bf16}
    \end{subfigure}
    ~ 
    \begin{subfigure}[t]{0.250\textwidth}
        \centering
        \includegraphics[width=1.0\linewidth]{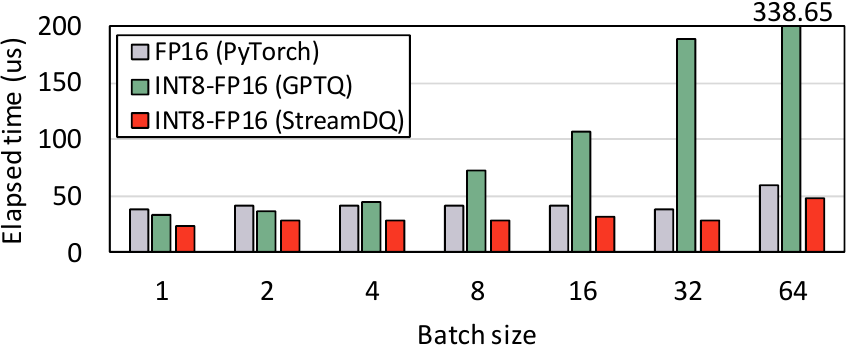}
        \caption{INT8-FP16}
        \label{fig:perf_int8-fp16}
    \end{subfigure}
    ~
        \begin{subfigure}[t]{0.250\textwidth}
        \centering
        \includegraphics[width=1.0\linewidth]{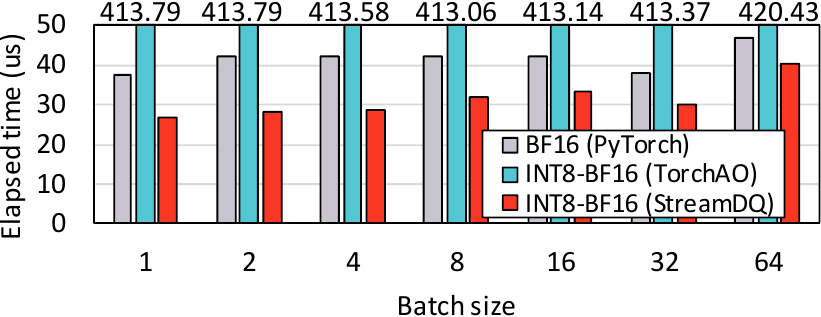}
        \caption{INT8-BF16}
        \label{fig:perf_int8-bf16}
    \end{subfigure}

    \begin{subfigure}[t]{0.250\textwidth}
        \centering
        \includegraphics[width=1.0\linewidth]{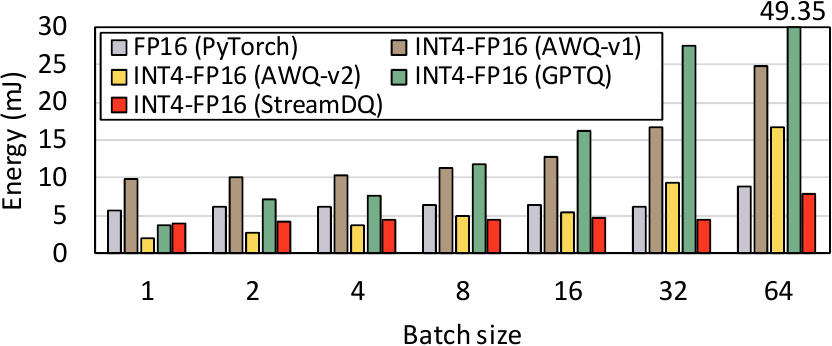}
        \caption{INT4-FP16}
        \label{fig:energy_int4-fp16}
    \end{subfigure}
    ~ 
    \begin{subfigure}[t]{0.250\textwidth}
        \centering
        \includegraphics[width=1.0\linewidth]{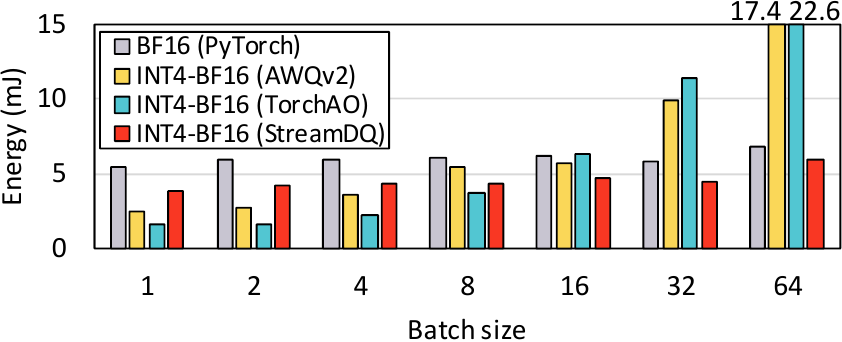}
        \caption{INT4-BF16}
        \label{fig:energy_int4-bf16}
    \end{subfigure}
    ~ 
    \begin{subfigure}[t]{0.250\textwidth}
        \centering
        \includegraphics[width=1.0\linewidth]{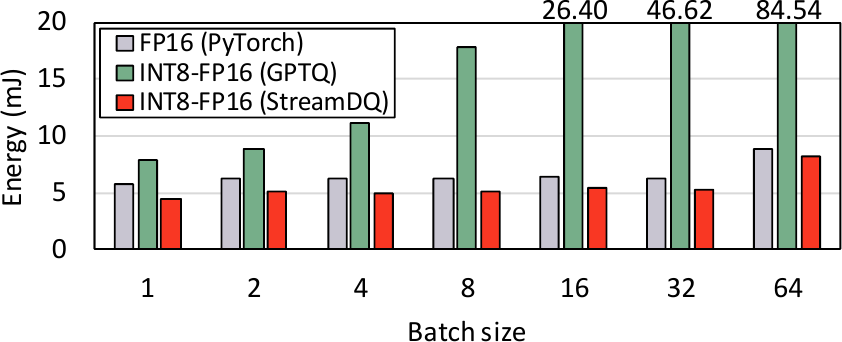}
        \caption{INT8-FP16}
        \label{fig:energy_int8-fp16}
    \end{subfigure}
    ~
        \begin{subfigure}[t]{0.250\textwidth}
        \centering
        \includegraphics[width=1.0\linewidth]{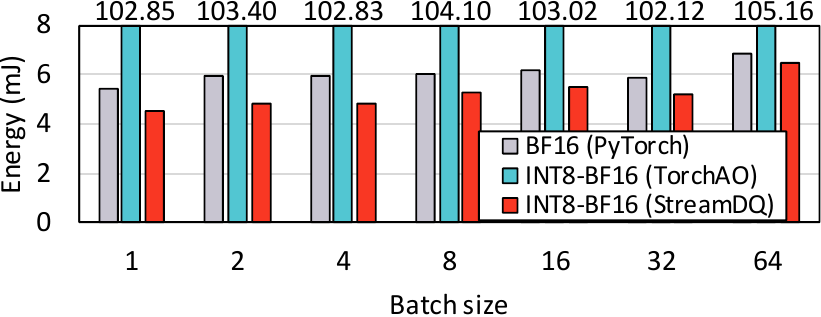}
        \caption{INT8-BF16}
        \label{fig:energy_int8-bf16}
    \end{subfigure}
\caption{
Runtime (a)--(d) and energy consumption (e)--(h) of mpGEMM (lower is better), comparing StreamDQ with software-based approaches.
Each experiment uses $K=N=4096$ and varies $M$ in the GEMM size ($M \times K \times N$).
Subfigure labels denote the quantized weight format and GEMM compute format.
}
\label{fig:eval_gemm_perf}
%\vspace{-1.2em}
\end{figure*}

\begin{figure*}[t]
    \centering
    \begin{subfigure}[t]{0.321\textwidth}
        \centering
        \includegraphics[width=1.0\linewidth]{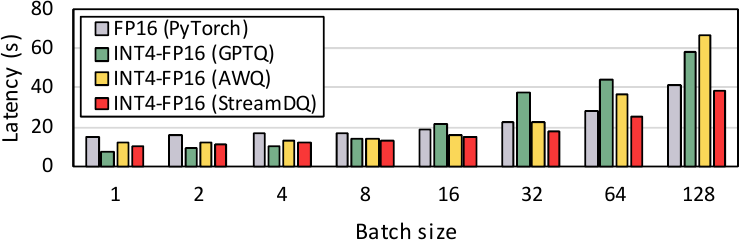}
        \caption{Llama-3.1-8B-Instruct (INT4)}
        \label{fig:lat_llama_int4}
    \end{subfigure}
    ~ 
    \begin{subfigure}[t]{0.321\textwidth}
        \centering
        \includegraphics[width=1.0\linewidth]{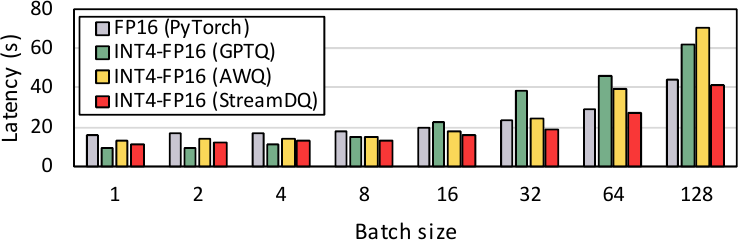}
        \caption{Qwen3-8B (INT4)}
        \label{fig:lat_qwen_int4}
    \end{subfigure}
    ~ 
    \begin{subfigure}[t]{0.321\textwidth}
        \centering
        \includegraphics[width=1.0\linewidth]{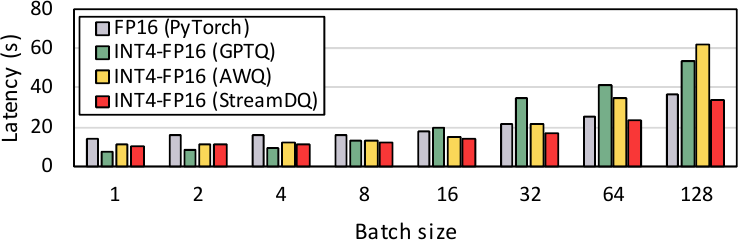}
        \caption{Mistral-7B-Instruct-v0.3 (INT4)}
        \label{fig:lat_mistral_int4}
    \end{subfigure}
    \\
    \begin{subfigure}[t]{0.321\textwidth}
        \centering
        \includegraphics[width=1.0\linewidth]{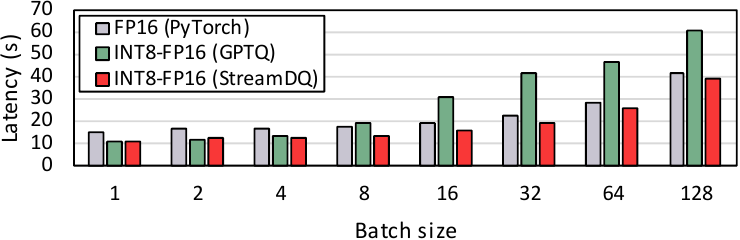}
        \caption{Llama-3.1-8B-Instruct (INT8)}
        \label{fig:lat_llama_int8}
    \end{subfigure}
    ~ 
    \begin{subfigure}[t]{0.321\textwidth}
        \centering
        \includegraphics[width=1.0\linewidth]{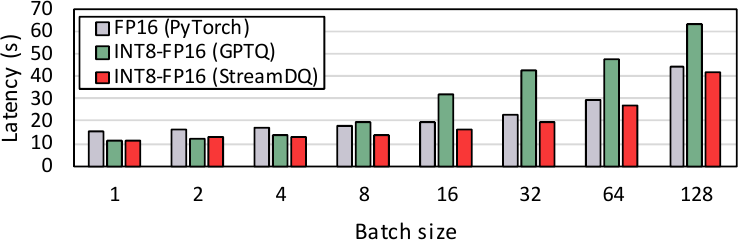}
        \caption{Qwen3-8B (INT8)}
        \label{fig:lat_qwen_int8}
    \end{subfigure}
    ~ 
    \begin{subfigure}[t]{0.321\textwidth}
        \centering
        \includegraphics[width=1.0\linewidth]{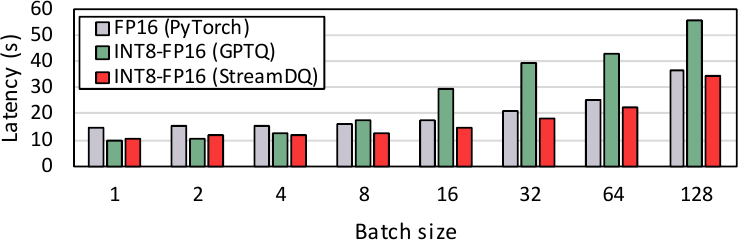}
        \caption{Mistral-7B-Instruct-v0.3 (INT8)}
        \label{fig:lat_mistral_int8}
    \end{subfigure}

    \begin{subfigure}[t]{0.321\textwidth}
        \centering
        \includegraphics[width=1.0\linewidth]{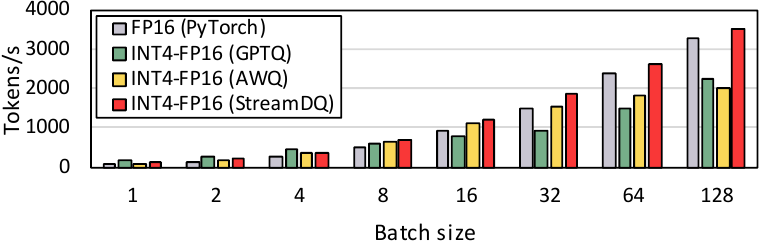}
        \caption{Llama-3.1-8B-Instruct (INT4)}
        \label{fig:tps_llama_int4}
    \end{subfigure}
    ~ 
    \begin{subfigure}[t]{0.321\textwidth}
        \centering
        \includegraphics[width=1.0\linewidth]{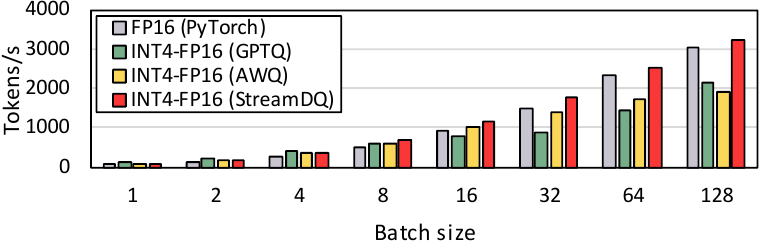}
        \caption{Qwen3-8B (INT4)}
        \label{fig:tps_qwen_int4}
    \end{subfigure}
    ~ 
    \begin{subfigure}[t]{0.321\textwidth}
        \centering
        \includegraphics[width=1.0\linewidth]{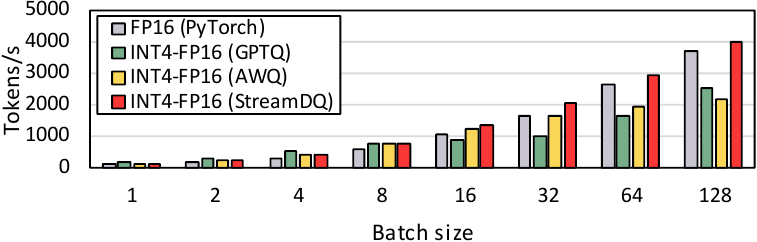}
        \caption{Mistral-7B-Instruct-v0.3 (INT4)}
        \label{fig:tps_mistral_int4}
    \end{subfigure}
    \\
    \begin{subfigure}[t]{0.321\textwidth}
        \centering
        \includegraphics[width=1.0\linewidth]{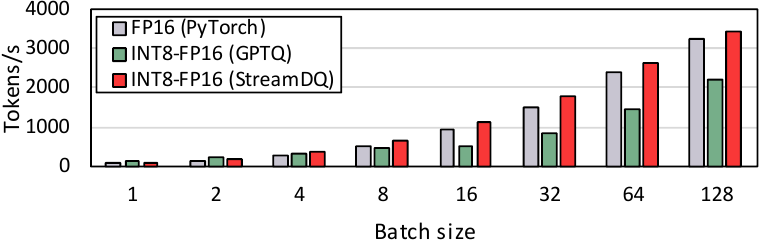}
        \caption{Llama-3.1-8B-Instruct (INT8)}
        \label{fig:tps_llama_int8}
    \end{subfigure}
    ~ 
    \begin{subfigure}[t]{0.321\textwidth}
        \centering
        \includegraphics[width=1.0\linewidth]{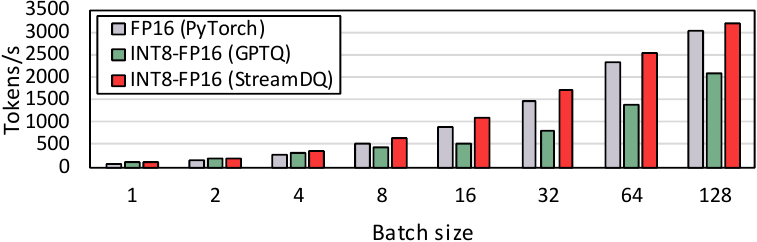}
        \caption{Qwen3-8B (INT8)}
        \label{fig:tps_qwen_int8}
    \end{subfigure}
    ~ 
    \begin{subfigure}[t]{0.321\textwidth}
        \centering
        \includegraphics[width=1.0\linewidth]{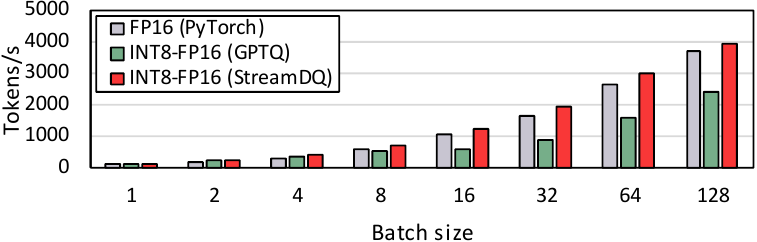}
        \caption{Mistral-7B-Instruct-v0.3 (INT8)}
        \label{fig:tps_mistral_int8}
    \end{subfigure}
    
\caption{
End-to-end inference latency (lower is better) in (a)--(f) and decode throughput (higher is better) in (g)--(l), comparing StreamDQ with software-based approaches.
All experiments use input and output sequence lengths of 1024.
The notation in each subplot denotes the model name and the data type of the quantized weight matrix.
}
\label{fig:eval_inf_lat}
    \vspace{-1.2em}
\end{figure*}

\subsection{Mixed-Precision GEMM Performance and Energy}
\label{sec:mixed_precision_gemm}
This section evaluates the performance and energy consumption of mpGEMM for StreamDQ and existing software-based approaches. Each experiment was repeated 16 times, and average values are reported. FP16 and BF16 baselines (PyTorch) represent full-precision GEMM (fpGEMM). All mpGEMM kernels use fused DQ-GEMM kernels, except TorchAO’s INT8$\rightarrow$BF16 implementation, which remains split because no fused version is available.

\textbf{Performance.}
As shown in Figs.~\ref{fig:eval_gemm_perf}-(a)--(d), StreamDQ consistently outperforms software-based approaches across batch sizes, achieving up to 7.08$\times$ speedup over fused-kernel approaches. 
For small batches (e.g., $\leq$ 4), AWQ-v2 slightly outperforms StreamDQ because fused DQ-GEMM kernels leverage shared memory to avoid extra HBM accesses in the memory-bound regime. As batch size increases (e.g., $\geq$ 8), GEMM execution becomes compute-bound, and StreamDQ outperforms all fused-kernel approaches, including AWQ-v2, because limited CUDA-core-based dequantization throughput restricts tensor-core utilization and degrades overall GEMM performance.

%As shown in Figs.~\ref{fig:eval_gemm_perf}-(a)--(d), StreamDQ consistently outperforms software-based approaches across batch sizes, achieving up to 7.08$\times$ speedup over fused-kernel baselines. For small batches (e.g., $\leq$ 4), AWQ-v2 slightly outperforms StreamDQ because fused DQ-GEMM kernels leverage shared memory to avoid extra HBM accesses. In these memory-bound settings, tensor-core utilization is not the primary bottleneck, allowing AWQ-v2 to achieve the highest performance among software methods.

%As batch size increases (e.g., $\geq$ 8), GEMM execution transitions into a compute-bound regime. In this regime, StreamDQ outperforms all fused-kernel approaches, including AWQ-v2. Fused kernels begin to suffer from the limited throughput of CUDA-core-based dequantization, which restricts tensor-core utilization and degrades overall GEMM performance. As a result, CUDA-core-based dequantization becomes the dominant bottleneck in software-based fused-kernel approaches.

For INT8$\rightarrow$FP16 workloads, StreamDQ achieves larger gains over GPTQ than in the INT4$\rightarrow$FP16 case because GPTQ incurs substantially higher memory traffic when loading the larger INT8 weight matrix.

Finally, Fig.~\ref{fig:eval_gemm_perf}-(d) shows the lower performance of TorchAO’s split DQ-GEMM kernel. 
Likewise, split kernels from other approaches, such as AWQ and GPTQ for the data types they support, also exhibit extremely low performance at small batch sizes and are therefore omitted from our evaluation. These kernels become beneficial only when GEMM is strongly compute-bound. We further discuss kernel selection as a function of batch size in Section~\ref{sec:eval_inf}.

%Finally, Fig.~\ref{fig:eval_gemm_perf}-(d) illustrates the lower performance of split DQ-GEMM kernel of TorchAO. Likewise, split kernels of other approaches such as AWQ and GPTQ for their supported data types also exhibit extremely low performance at small batch sizes while the results are excluded in this paper. These kernels benefit only when GEMM is strongly compute-bound. We discuss the kernel selection depending on batch sizes in Section~\ref{sec:eval_inf}

\textbf{Energy Consumption.}
Fig.~\ref{fig:eval_gemm_perf}-(e)--(h) reports energy consumption based on power data obtained from StreamDQ-Sim. 
The additional power introduced by the DQB is negligible, and its average power remains nearly constant across batch sizes. 
In contrast, software-based fused kernels incur increasing on-chip traffic and longer execution times as batch size grows, increasing both power draw and total energy consumption. As a result, StreamDQ reduces total energy consumption by up to 90.23\% relative to fused-kernel approaches across the evaluated configurations.
%As a result, StreamDQ delivers substantial energy savings across all evaluated configurations, reducing total energy consumption by up to 90.23\% relative to fused-kernel approaches.

\begin{figure*}[t]
    \centering
    \begin{subfigure}[t]{0.321\textwidth}
        \centering
        \includegraphics[width=1.0\linewidth]{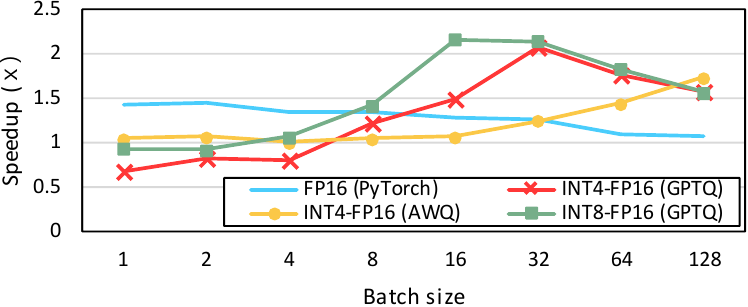}
        \caption{Llama-3.1-8B-Instruct}
        \label{fig:su_llama}
    \end{subfigure}
    ~ 
    \begin{subfigure}[t]{0.321\textwidth}
        \centering
        \includegraphics[width=1.0\linewidth]{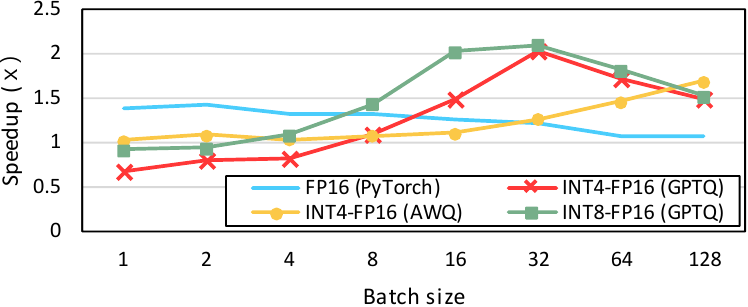}
        \caption{Qwen3-8B}
        \label{fig:su_qwen}
    \end{subfigure}
    ~ 
    \begin{subfigure}[t]{0.321\textwidth}
        \centering
        \includegraphics[width=1.0\linewidth]{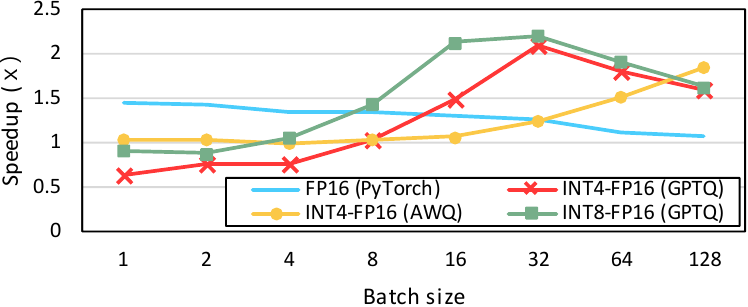}
        \caption{Mistral-7B-Instruct-v0.3}
        \label{fig:su_mistral}
    \end{subfigure}
\caption{
Speedup of StreamDQ over software-based approaches based on decode throughput, with input and output sequence lengths of 1024.
%The notation in each subplot denotes the model name and the data type of the quantized weight matrix.
}    
%    \caption{Speedups of StreamDQ over existing approaches based on the decode throughput where input and output sequence lengths are 1024. Notation for each experiment indicates model name and data type of quantized weight matrix.}
    \vspace{-1.2em}
    \label{fig:eval_inf_su}
\end{figure*}

\subsection{End-to-End LLM Inference Performance}\label{sec:eval_inf}
\textbf{Kernel Selection.}  
As discussed earlier, the performance of mpGEMM depends heavily on whether the workload is memory-bound or compute-bound. Selecting between fused and split DQ-GEMM kernels is therefore essential for maximizing performance. For fairness, we report the best GPTQ and AWQ performance at each batch size. We performed a heuristic search to identify when GPTQ and AWQ should transition between fused and split kernels. For GPTQ, the fused kernel is optimal for batch sizes $\leq$ 32, whereas the split kernel becomes more efficient for batches $\geq$ 64. Accordingly, we set the switching threshold to 64 in vLLM. AWQ transitions at batch sizes $\geq$ 256; since our evaluation covers up to batch size 128, AWQ operates entirely in fused mode (AWQ-v1).

\textbf{Performance.}  
Figs.~\ref{fig:eval_inf_lat}-(a)--(l) show the end-to-end inference latency and decode throughput results, while Fig.~\ref{fig:eval_inf_su} shows the corresponding speedups of StreamDQ over software-based approaches. StreamDQ consistently outperforms all baselines, achieving up to 54.68\% lower latency and 2.20$\times$ higher decode throughput. These gains primarily arise from eliminating software-level dequantization, while the small per-access latency added by the DQB is effectively amortized within the GEMM execution pipeline.

For small batch sizes (e.g., $\leq$ 4) in the INT4$\rightarrow$FP16 setting, GPTQ slightly outperforms StreamDQ because its fused kernel is advantageous in memory-bound regimes. As batch size increases and the workload becomes compute-bound, StreamDQ surpasses GPTQ. Beyond batch size 32, GPTQ switches to a split kernel to restore tensor-core utilization, but this introduces additional HBM write-back and reload overhead. In contrast, StreamDQ performs inline dequantization during memory loads, eliminating redundant data movement and maintaining high efficiency even under compute-bound conditions.

StreamDQ also consistently outperforms AWQ across most batch sizes, with the largest gap appearing at batch size 128. The gap is expected to narrow beyond batch size 256 when AWQ transitions to split-kernel execution, but within our evaluation range AWQ remains fused and increasingly constrained by CUDA-core dequantization overhead.

For INT8$\rightarrow$FP16 workloads, StreamDQ outperforms GPTQ from batch size 4 onward, with larger relative gains than in the INT4$\rightarrow$FP16 case because GPTQ incurs higher memory traffic when loading INT8 weights. 
The relative performance benefit of eliminating dequantization overhead naturally decreases at larger batch sizes, as the dequantization portion of total runtime decreases (see Fig.~\ref{fig:dq-overhead}).
While GPTQ and AWQ degrade noticeably in compute-bound regimes, StreamDQ maintains stable performance, demonstrating its strong suitability for high-throughput LLM inference.

% Daegun Yoon: 분량 상 일단 삭제 (2025/11/12)
% Daegun Yoon: 실험 결과로부터 유추 가능한 limitations인데 명시적으로 논문에 넣었다가 공격받을 수도 있으니 삭제해도 무방 (2025/11/10)
%\textbf{Limitations.} As we discussed earlier, StreamDQ gets more powerful for DQ-GEMM if the workload gets more compute-intensive. On the other hand, the performance of StreamDQ is bit limited compared to those of several fused kernel approaches for small batch sizes (e.g., $\leq$ 4). If a weight matrix gets larger in such memory-bound regimes, the performance of StreamDQ gets more limited than the fused kernels. For example, although StreamDQ outperforms GPTQ for every batch size in INT8-FP16 GEMM evaluation, StreamDQ shows worse performance than GPTQ for batch sizes one and two in inference evaluation. This is because weight matrices of particular linear layers within an LLM are larger than that in GEMM evaluation. Nevertheless, this disadvantage is limited to such memory-bound regimes, as the evaluation results revealed that StreamDQ outperforms existing approaches in most of cases.

\section{Related Works}

Efforts to accelerate mixed-precision GEMM (mpGEMM) for weight-only quantized models on conventional GPUs can be broadly divided into two categories: direct mpGEMM, which multiplies quantized weights and full-precision activations without explicit dequantization, and indirect mpGEMM, which first dequantizes weights to match the activation precision before computation.

\textbf{Direct mpGEMM.}
These approaches operate on quantized weights in their native formats. Integer-weight-based designs align floating-point activations for integer arithmetic~\cite{kim2023winning,jang2024figna,fang2025anda}; for example, iFPU~\cite{kim2023winning} and FIGNA~\cite{jang2024figna} pre-align activations for integer multiply-accumulate operations, while AnDa~\cite{fang2025anda} introduces a compact activation format. Other designs replace multipliers with approximate or LUT-based operators: BitMod~\cite{chen2025bitmod} and MixPE~\cite{zhang2024mixpe} decompose low-bit weights into binary components, AxCore~\cite{zou2025axcore} uses addition-based approximations, and LUT-based mpGEMM~\cite{park2022lut,mo2025lut,park2025figlut,wei2025t} precomputes partial products for activation-weight combinations. Although direct mpGEMM can improve computational efficiency and reduce storage demands, it generally requires precision-specific hardware and often disruptive changes to existing hardware, limiting flexibility and scalability across data formats.

\textbf{Indirect mpGEMM.}
These methods reuse existing GPU GEMM units by dequantizing low-precision weights to match the activation precision before computation~\cite{gptq,lin2024awq,kim2024quick}. This avoids custom hardware and supports arbitrary precision pairs, but the dequantization step incurs significant compute and memory traffic overhead. Although recent work mitigates shared-memory bank conflicts~\cite{kim2024quick}, dequantized weights can still spill to HBM in large-batch LLM inference because of limited on-chip storage, leading to substantial performance degradation.

\textbf{StreamDQ.} StreamDQ addresses these limitations by performing on-the-fly dequantization in the HBM base die. It preserves compatibility with existing GPU GEMM pipelines while avoiding CUDA-core-based dequantization overhead and redundant HBM write-back and reload of dequantized weights. It also provides efficient support for multiple quantized-to-dequantized format pairs in high-throughput LLM inference.

\section{Conclusion}
We present StreamDQ, a near-memory dequantization architecture for high-throughput LLM inference that performs on-the-fly dequantization in the HBM base die while preserving conventional GPU load semantics. By offloading dequantization from CUDA cores and using lightweight sideband tagging for control, StreamDQ reduces GPU-side instruction overhead, on-chip traffic, and redundant off-chip traffic in compute-bound regimes, while supporting multiple quantized-to-dequantized format pairs within the power, area, and thermal constraints of the HBM base die. Our evaluation shows substantial improvements in inference latency and throughput with low area and power overhead.

% \begin{acks}
%    This document is derived from previous conferences, in particular MICRO 2013, ASPLOS 2015, MICRO 2015-2025, ISCA 2025, as well as SIGARCH/TCCA's Recommended Best Practices for the Conference Reviewing Process. 
% \end{acks}

%%%%%%% -- PAPER CONTENT ENDS -- %%%%%%%%

\bibliographystyle{ACM-Reference-Format}
\bibliography{sample-base}

\end{document}